\documentclass[10pt,preprint2]{aastex}
\usepackage{amsmath}
\usepackage[hypertex]{hyperref}
\usepackage{epsfig}
\usepackage{pdflscape}
\usepackage{natbib}
\usepackage{verbatim}
\usepackage[margin=2.5cm]{geometry}

\newcommand{\planck}{{\it Planck\/}}
\newcommand{\wmap}{{\it WMAP\/}}
\newcommand{\xmm}{{\it XMM-Newton\/}}

\newcommand{\mpch}{{\rm \ Mpc\ h^{-1}}}
\newcommand{\hfi}{{\em HFI}}
\newcommand{\lfi}{{\em LFI}}
\newcommand{\osb}{{OMCP}}

\begin{document}
 
\title{Measuring Bulk Flow of Galaxy Clusters using Kinematic Sunyaev-Zel'dovich effect: Prediction for Planck}
\author{D.S.Y. Mak, E. Pierpaoli}
\affil{University of Southern California, Los Angeles, CA, 90089--0484}
\author{S.J. Osborne}
\affil{Stanford University, 382 Via Pueblo, Varian Building, Stanford, CA 94305}

\begin{abstract}
We predict the performance of the \planck\  satellite  in determining  the bulk flow through kinetic Sunyaev-Zeldovich (kSZ) measurements.
As velocity tracers, we use ROSAT All-Sky Survey (RASS) clusters as well as  expected cluster catalogs  from the upcoming  missions  \planck\  and eRosita (All-Sky Survey: EASS). 
We implement a semi-analytical approach to simulate realistic \planck\ maps   as well as   \planck\  and eRosita cluster catalogs. We adopt an unbiased kinetic SZ filter (UF) and matched filter (MF) to maximize the cluster kSZ signal to noise ratio. 
We find that the use of  \planck\  CMB maps  in conjunction  with the currently existing ROSAT cluster  sample improves current upper limits on the bulk flow determination by a factor $\sim5$ ($\sim10$) when using the MF (UF).
The accuracy of bulk flow measurement increases with the depth and abundance of the cluster sample: for an input  bulk velocity of 500 km/s, the UF recovered velocity errors decrease from 94 km/s for RASS, to 73 km/s for \planck\ and  to 24 km/s for EASS; while the systematic bias decreases from 44\% for RASS,  5\% for \planck, to 0\% for EASS.
 The $95\%$ upper limit for the recovered bulk flow direction  $\Delta\alpha$ ranges between $4^{\circ} ~\rm{and} ~ 60 ^{\circ}$ depending on cluster sample and adopted filter.
 The kSZ dipole determination is mainly limited by  the effects of thermal SZ  (tSZ)
 emission in all cases but the one of EASS clusters analyzed with the unbiased filter.  
 This fact makes the UF preferable to the MF when analyzing \planck\ maps.

\end{abstract}
\keywords{Cosmology: cosmic microwave background, observations, diffuse radiation}

\section{Introduction}

Peculiar velocities, along with inhomogeneities, can be  used to constrain cosmology.
A coherent, large scale peculiar velocity, also called bulk flow, may originate from spatial inhomogeneities in the mass distribution of large scale structures around us. The standard inflationary model predicts that the rms bulk velocity within a sphere of radius R decreases linearly with comoving distance in the $\Lambda$CDM universe with $V_{\rm rms}(r>50-100\mpch)\approx250(\frac{100 \mpch}{r})$ km/s~\citep{Kashlinsky1991}. Galaxy cluster peculiar velocity surveys at scales $R\le60 \mpch$ generally agree with theoretical predictions of the cluster bulk velocities. However, recent measurements at larger scales ($R\ge100 \mpch $) indicate that the bulk flow velocity is significantly larger than the $\Lambda$CDM prediction with statistical significance up to $3\sigma$~\citep{Kashlinsky2008,Feldman2010}. 

\begin{figure*}
  \begin{center}
    \leavevmode
       \includegraphics[width=90mm, angle=90]{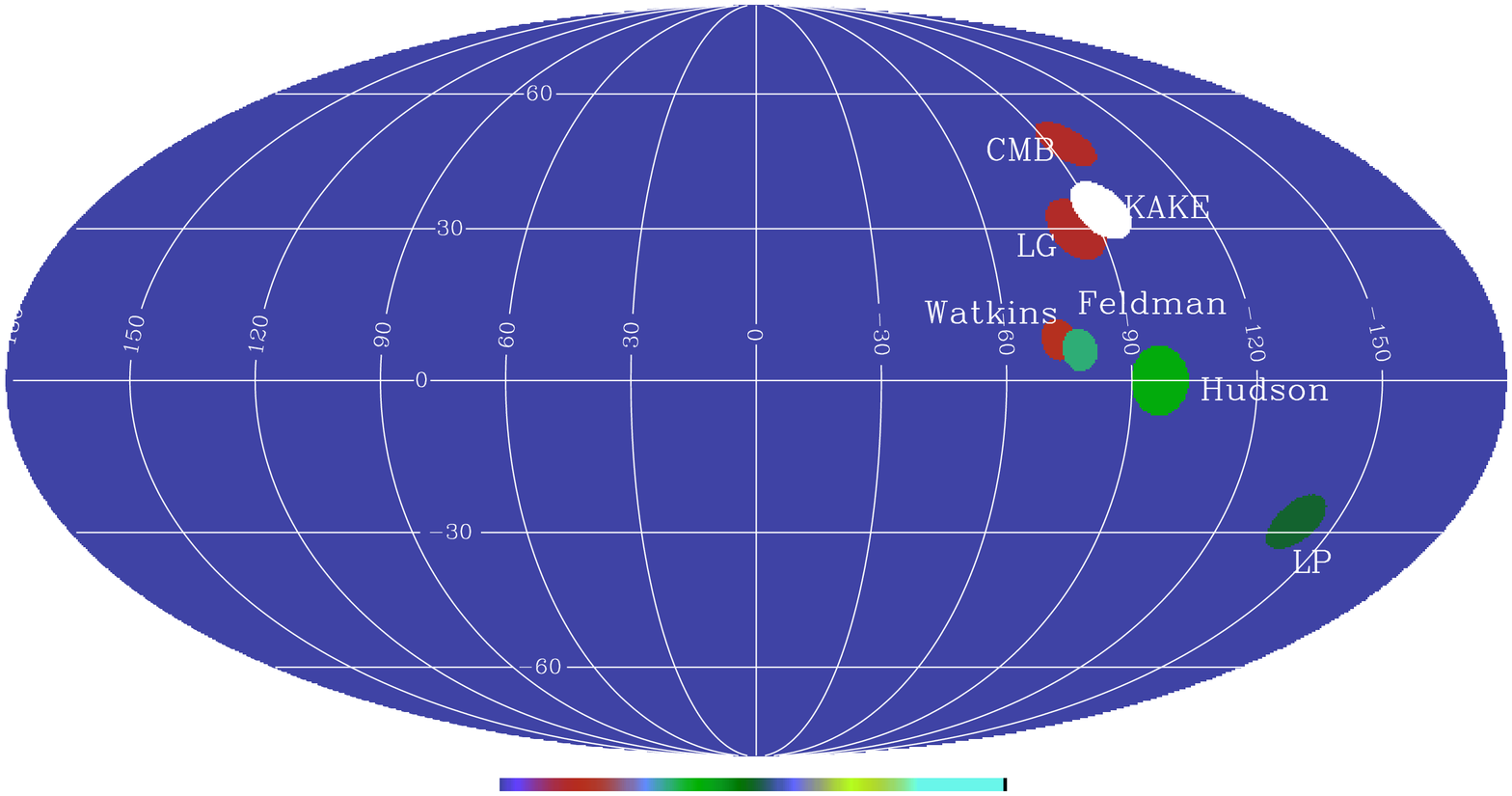}
       \caption{The dipole direction in Galactic coordinates of current bulk flow measurements.  (a) CMB $b=48.26\pm0.03^{\circ}$, $l=263.99\pm0.14^{\circ}$~\citep{Jarosik2010}; (b) Local Group $b=30\pm3^{\circ}$, $l=276\pm3^{\circ}$~\citep{Kogut1993}; (c) KAKE $b=34^{\circ}$, $l=267^{\circ}$; (d)~\citet{Watkins2009} $b=8\pm6^{\circ}$, $l=287\pm9^{\circ}$; (e)~\citet{Lauer1994} $b=-28\pm27^{\circ}$, $l=220\pm27^{\circ}$; (f)~\citet{Hudson2004} $b=0\pm11^{\circ}$, $l=263\pm13^{\circ}$; (f)~\citet{Feldman2010} $b=6\pm6^{\circ}$, $l=282\pm11^{\circ}$. The size of the colored region is proportional to the amplitude of the measured bulk flow. The color code represents the convergence depth of the bulk flow, in units of $\mpch $. }
     \label{f:dipole}
  \end{center}
\end{figure*}

At scales $R\le60 \mpch $, an enhancement of the bulk flow with respect to the predicted $\Lambda$CDM value in the local universe is attributed to a large scale void or overdensity at these depth. This is thought to be the cause of the Local Group (LG) motion with respect to the CMB rest frame, with velocity $v=627\pm22$ km/s towards $l=276^{\circ}$, $b=+30^{\circ}$, in alignment with the CMB dipole~\citep{Kogut1993}. The measured value is within the cosmic variance limit.~\citet{Dressler1987} and~\citet{Lynden1988} identified the Great Attractor (GA), a mass concentration of $\sim10^{16}M_{\odot}$, as the origin of the flow at $<60 \mpch $. On large scales where $\Lambda$CDM predicts bulk flows with negligible amplitude, observations show the contrary; ~\citet{Lauer1994} found a strong bulk flow signature with $v=561\pm284$ km/s towards $l=220^{\circ}$, $b=-28^{\circ}$ ($\pm27^{\circ}$) at a depth of 110 $\mpch$. Using a sample of 119 Abell clusters within $150 \mpch $ they found that the flow originate from a mass concentration beyond $100 \mpch $. Similarly,~\citet{Feldman2010} found a bulk flow on scales of $\sim 100 \mpch $ with $v=416\pm78$ km/s towards $l=282^{\circ}\pm11^{\circ}$, $b=+6^{\circ}\pm6^{\circ}$, in disagreement with the WMAP5 cosmological parameters at 99.5\% confidence. The direction and scale of these bulk flows are shown in Figure~\ref{f:dipole}.

Several theories have been suggested to explain the high bulk flow velocities at large scales. One explanation is that pre-inflationary fluctuations in scalar fields on superhorizon scales gives a titled universe (\citealt{Turner1991},~\citealt{Kashlinsky1994},~\citealt{Mersini2009}). In this picture, matter slides from one side of our Hubble volume to the other, producing an intrinsic CMB dipole anisotropy as seen in the matter rest frame. This inhomogeneity generates a bulk flow with correlation length of order the horizon size. Alternatively,~\citet{Wyman2010} have showed that a strengthened gravitational attraction at late times can speed up structure formation and increase peculiar velocities. Using N-body simulations, they found an enhancement in large scale, $R>100 \mpch $, bulk flow velocities of up to $\sim40\%$ relative to the $\Lambda$CDM cosmology. A similar approach using modified gravity is discussed in~\citet{Afshordi2009} and~\citet{Khoury2009}.

\citet{Kashlinsky2000} have proposed a method aimed at determining the largest scale bulk flows from galaxy cluster peculiar velocities measured using the kinetic Sunyaev-Zeldovich (kSZ) effect. If many galaxy clusters are moving with a coherent motion with respect to the CMB rest frame, the kinematic part of the SZ signal acquires a dipole moment. Since the kSZ signal is proportional to line of sight velocity, such a measurement directly probes the bulk flow, free of distance measurement errors. Several authors attempted to measure the bulk flow using the measured kSZ effect in \wmap\ data. \cite{Kashlinsky2008} (hereafter KAKE, and later~\citealt{Kashlinsky2010}) first utilized this method, claiming a large-scale flow with $v>600$ km/s out to $\sim575 \mpch $, without sign of convergence to the $\Lambda$CDM predicted value. However, by repeating the same method~\citet{Keisler2009} did not detect a statistically significant bulk flow.~\citet{Osborne2010} (hereafter \osb), used filters constructed to enhance the signal to noise of the kinetic signal and found no significant velocity dipole in the WMAP 7 year data.  More specifically, they found a 95\% bulk flow upper limits of the order of 4600 km/s in the direction of the KAKE They also  showed that the matched filter outperforms the unbiased one when  \wmap\ data are used, and demonstrated that CMB and instrument noise dominate the uncertainties.

In this work, we apply the scheme of \osb\ to study the capability of \planck\ data and future cluster surveys to measure bulk flows. The use of Planck maps is expected to produce improved results with respect to the  WMAP case because of reduced instrument noise, wider frequency coverage (ensuring better foregrounds' subtraction) and increased spatial resolution of this mission. In addition,  ~\planck\ will  also produce the first all-sky SZ survey with a median redshift of $z=0.3$ (\planck\ Blue Book) and will ensure a better performance   in  separating the  tSZ from the kSZ signal for any cluster sample considered.

We investigate to what extent the use of Planck maps, in combination with data for existing ROSAT clusters, improves on bulk flow determination from WMAP. We also assess the expected performances of bulk flow measurements for upcoming all--sky cluster catalogs, such as the ones derived from \planck\ and eRosita satellites. Such samples are more abundant and extend to higher redshifts than the one in hand.

Our goals are: (i) to determine the sensitivity of the cluster velocity dipole measurement with the \planck\ specifications and assess the nature of the uncertainty; (ii) to study which cluster survey can best constrain the bulk flow; (iii) to study the performance of the filters used in \osb\ with the \planck\ setup.

This paper is organized as follows. The bulk flow velocity expected from the $\Lambda$CDM model is calculated in section~\ref{s:theory}. In section~\ref{s:outline}, we briefly describe the procedure we use to measure the bulk flow velocity. In sections~\ref{s:cat} and~\ref{s:szmap} we give details of the SZ and X-ray cluster catalogs we use and describe the procedure we adopt to generate simulated SZ maps. In section~\ref{s:filter}, we present the two filters we use to reconstruct the kSZ signal from the CMB maps. In section~\ref{s:dipole}, we describe the analysis pipeline we use to measure and calibrate the cluster dipole.
In section~\ref{s:error}, we describe the systematic effects that may contaminate our results. The results are presented in section~\ref{s:results}, followed by our conclusions in section~\ref{s:con}. Throughout this paper, we assume a $\Lambda$CDM cosmological model with $\Omega_m=0.3$, $\Omega_\Lambda=0.7$, $h=0.72$, $w=-1$, $\sigma_8=0.8$.

\section{Theoretical Aspects}
\label{s:theory}
\subsection{The Kinetic SZ effect as a probe of velocity}
\label{ss:thsz}

The SZ effect~\citep{Sunyaev1970} is a secondary CMB anisotropy caused by the scattering of CMB photons by high energy electrons, such as those inside the intra-cluster medium. The fractional change in temperature of the CMB photons is the sum of two componets: $\delta T= \Delta T_{\rm kSZ} + \Delta T_{\rm tSZ}$, where the first term and second terms are the kinetic  (kSZ) and termal (tSZ) Sunyaev-Zeldovich effect respectively. The kSZ is caused by the Doppler shifting of CMB photons due to the peculiar motion of the galaxy cluster with respect to the CMB rest frame. The fractional temperature change is:

\begin{equation}
\left ( \frac{\Delta T}{T_{\rm CMB}} \right )_{\rm kSZ}= - \frac{v_p}{c} \tau
 \label{eq:ksz}
 \end{equation}
 
 \noindent where $v_p$ is the line-of-sight peculiar velocity of the cluster, $\tau=\sigma_{\rm T} \int n_{\rm e} dl$ is the optical depth of the cluster, $n_e$ is the electron density, and $\sigma_T$ is the Thompson scattering cross section. For typical galaxy clusters with $\tau\sim10^{-3}$ and $v\sim500$ km/s, we expect a CMB temperature change of $\Delta T_{\rm kSZ}\sim 5\mu K$ at the location of a galaxy cluster. To date, the kSZ effect has not been measured in individual galaxy cluster. Nevertheless, if many galaxy clusters are moving in the same direction, their velocity field creates a dipolar pattern in the CMB radiation at large scale and leads to a net dipole moment $C_{1,ksz}=T_{\rm CMB}^2\left \langle \tau \right \rangle^2 V_{\rm bulk}^2/c^2$ (KAKE), where $\left \langle \tau \right \rangle$ is the mean optical depth of the cluster sample.

The thermal Sunyaev-Zeldovich (tSZ) effect is the boosting in energy of CMB photons scattered by hot electrons in the intra cluster medium and results in a fractional temperature change of:
 
\begin{equation}
\left ( \frac{\Delta T}{T_{\rm CMB}} \right )_{\rm tSZ}=y f(x) 
\label{eq:tsz}
\end{equation}
\noindent where $y=\sigma_{\rm T}\int dl (k_{\rm B}T_e)/(m_e c^2) n_e$ is the Comptonization parameter, $T_e$ is the electron temperature of the cluster, $f(x)=x(e^x+1)/(e^x-1)-4$ is the frequency dependence of the tSZ effect and $x=h\nu/k_{\rm B}T_{\rm CMB}$. The minimum and maximum temperature changes occur at $\nu=143$ GHz and $\nu=353$ GHz with a null at $\nu=217$ GHz. For typical clusters, $T_e\sim10$ keV gives $\Delta T_{\rm tSZ}\sim100\mu K$. In individual clusters, the tSZ effect is typically larger than the kSZ effect:

\begin{equation}
\frac{\Delta T_{\rm kSZ}}{\Delta T_{\rm tSZ}}\sim 0.1f(x)\left (\frac{v_p}{300\ {\rm km/s}} \right )  \left( \frac{T_e}{5 \rm keV} \right )^{-1}
\end{equation}

\noindent Nevertheless, the dipole component of the kSZ effect may dominate over the statistical dipole component of the tSZ effect if a bulk flow is present. Furthermore, the different frequency dependence of the two effects allows us to extract the kSZ signal.

\subsection{Expected Bulk Flow Velocity from $\Lambda$CDM model}
\label{ss:pre}
In the linear theory of structure formation the peculiar velocity field of galaxy clusters is related to the matter overdensity through the continuity equation:

\begin{equation}
\vec{v_k}=if(z)\frac{\delta_k}{k}\hat{k}
\label{eq:vk}
\end{equation}
\noindent where $f(z)\equiv\frac{a}{D_{1}}\frac{dD_{1}}{da}=H(z)\left|\frac{dD_{1}(z)}{dz}\right| $, and $D_{1}$ is the growth factor (equation~\ref{eq:growth}).
 
At scales much larger than the size of a cluster, the peculiar velocity field is correlated within a given region leading to a coherent motion, or bulk flow, of the objects inside the region. Assuming the distribution of clusters is isotropic, the bulk flow velocity within a region of size $R$ is given by:

\begin{equation}
\sigma^2_{v}(z)=f^2(z)\int{dk\  \frac{P_m(k)}{2\pi^2}|W(kR)|^2}
\label{eq:sigv}
\end{equation}

\noindent where $W(kR)$ is the Fourier transform of the top-hat window function, $R=\int_0^z{c/H(z') dz'}$ is the comoving radius and $P_m(k)$ is the present-day linear matter power spectrum. The cosmology dependence is embedded in the matter power spectrum $P_m(k)$ and its time evolution. In particular the rms velocity, $\sigma_v$, is proportional to the amplitude of the matter power spectrum $\sigma_8$ which is known with  an  uncertainty $<10\%$ (e.g.~\citealt{Larson2010}).

We incorporate the effect of the different redshift distributions of the clusters in the catalogs we use by weighting the bulk flow velocity by a selection function $\phi(z)$. The selection function takes into account the fact that the survey is not complete at all of the redshifts we use. Then,

\begin{equation}
\sigma^2_{v}(z)=\int{dk \  \frac{P_m(k)}{2\pi^2}|W(kR)|^2} \left (\int_0^z{ \phi(z') f(z') \frac{dV}{dz'} dz'} \right )^2
\label{eq:sigv2}
\end{equation}

\noindent where $V(z)$ is the comoving volume, $\phi(z)$ is the comoving number density or selection function. We calculate $\phi(z)$ from the halo mass function (described in appendix~\ref{ss:numcount}) and cluster mass limits for each survey (section~\ref{s:cat}),

\begin{equation}
\phi(z)=\bar{n}(z) = \int_{M_{\rm min}(z)}^{\infty} dM  \frac{dn(M,z)}{dM}
\end{equation}

\noindent where $M_{\rm min}(z)$ is the limiting mass of object in the cluster survey at redshift z. $\phi(z)$ is normalized such that

\begin{equation}
 \int_0^z \phi(z') (dV(z')/dz'd\Omega) dz' =1 \nonumber
 \end{equation} 

\begin{figure*}
  \begin{center}
   \leavevmode
       \includegraphics[width=90mm]{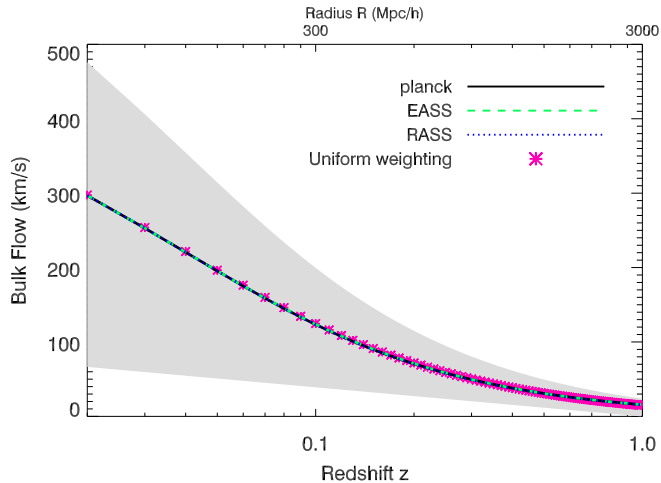}
       \caption{The theoretical bulk flow velocity in $\Lambda$CDM cosmology, using selection functions for the three cluster surveys, smoothed over a top hat window function $W(kR)$ with the comoving sphere centered on the observer with radius R. Also plotted is the amplitude of the     expected bulk flow with uniform selection function. The shaded region is the uncertainty of the expected bulk flow from sample variance.}
     \label{f:rmsv}
  \end{center}
\end{figure*}

Figure~\ref{f:rmsv} shows the rms bulk flow velocity predicted by equation~\ref{eq:sigv2} for the three cluster samples considered in this work. The effect of the selection function is small that the velocities among the three cluster samples are no different than $1\%$ at all redshifts. For comparison, the rms bulk flow computed with no selection function is also plotted. The sample variance of the velocity distribution is the largest source of uncertainty. For a Gaussian density field the amplitude of the bulk flow has a Maxwellian distribution with the probability of having a velocity lying between $V$ and $V+dV$ given by $P(V)dV\propto V^2\exp(-1.5V^2/\sigma_v^2)dV$. The 95\% confidence limits on the measured bulk flow of amplitude $V$ are then $V/3<V<1.6V$. This is the shaded region in the plot. 

In addition to large scale correlated cluster velocities, each cluster has a peculiar velocity caused by matter inhomogeniety on cluster scale ($R\sim8 \mpch $). We include this components in our simulations. 

\section{Methodology}
\label{s:outline}
\subsection{Assumptions and Definitions}
In this section we define  the different physical processes contributing to the sky signal, as well as the \planck\ instrumentation. 

The observed dipole signal at clusters' locations has, in principle, several contributions: 
\begin{equation}
a_{1m}=a_{1m}^{\rm CMB} + a_{1m}^{\rm noise} +a_{1m}^{\rm tSZ} + a_{1m}^{\rm kSZ} + a_{1m}^{\rm point} + a_{1m}^{\rm Gal} 
\label{eq:alms}
\end{equation}
\noindent where $a_{1m}^{\rm i}$ are the dipole coefficients of the CMB, instrument noise, thermal SZ, kinetic SZ, extragalactic radio and infrared point sources, and galactic signal. An appropriate temperature map mask  can be used to suppress the galactic components and fit for the cluster  velocity dipole outside the galactic region. We therefore do not consider the effect of galactic foregrounds in this work and apply a cut at $|b|\le7^\circ$ to the simulated \planck\  maps. The contribution from infrared point sources is not well characterized  at the present time and no model is available to accurately describe their abundances within galaxy clusters. We therefore do not consider the infrared point source signal in this work, and only simulate the first 4 terms in equation~\ref{eq:alms} as well as the radio point source term.

Our goal is to separate $a_{1m}^{\rm kSZ}$ from the other components and calculate its amplitude in \planck\ maps using different cluster samples. The detector noise levels and beam sizes we use are summarized in Table~\ref{t:table_planck_channel}. Although CMB observations by \planck\ cover 9 frequencies from 30 GHz to 857 GHz, we only consider the 6 channels between 44 GHz and 353 GHz because of large potential foreground contamination outside this range.

\begin{deluxetable}{lrrrrrr}
\tablecaption{Characteristics of the \planck\ channels}
\tablewidth{0pc}
\tablehead{
\colhead{\planck~ Channel} &\colhead{1}& \colhead{2}&\colhead{3}&\colhead{4}&\colhead{5}&\colhead{6} 
}
\startdata
Center frequency $\nu_0$	 (GHz)			
&44  & 70   & 100 & 143 & 217 & 353		\\
Resolution $\Delta\theta$ (FWHM)  & $26\farcm8$ & $13\farcm1$ & $9\farcm2$ &$7\farcm1$ & $5\farcm0$ & $5\farcm0$ \\
Noise level $\sigma_\mathrm{N}$ ($\frac{\Delta T}{T_{\rm CMB}}$, $10^{-5}$)	& 1.1 & 2.2 & 0.6 & 1.0 & 1.6 & 4.9 \\
\enddata
\tablecomments{Characteristics of the \planck\ \lfi-receivers (column 2-3) and \hfi-bolometers (column 4-7): center frequency $\nu_0$, angular resolution 
$\Delta\theta$ in FWHM, and instrument noise variance per pixel $\sigma_\mathrm{N}$ (thermodynamic temperature units).}
\label{t:table_planck_channel}
\end{deluxetable}

\subsection{Outline of the Method}
The steps we take to estimate the cluster kinetic SZ dipole velocity are:
\begin{enumerate}
\item We simulate cluster catalogs using halo model and cluster self similar scaling relations. We do so for three different cluster samples from the ROSAT All-Sky Survey, \planck, and eRosita All-Sky Survey.
\item  Full-sky maps of the simulated kinetic and thermal SZ are created in the 6 \planck\ frequency channels together with realizations of the CMB. These maps are convolved with Gaussian beams and detector noise is added using the properties given in Table~\ref{t:table_planck_channel}. 
\item The maps are filtered to enhance the kinetic SZ signal while suppressing the CMB and instrument noise terms and removing the thermal SZ signal. 
\item The filtered maps are used to calculate the dipole moment of the reconstructed kSZ signal. The dipole is calculated at the cluster positions using different redshift shells.
\item Temperature dipoles (in units of Kelvin) are converted into velocity dipoles (in units of km/s) by a calibration matrix $\mathbf{M}$, which is calculated from kSZ realizations with pre-defined bulk flow amplitudes, i.e. $\mathbf{a}_{\rm V}=\mathbf{Ma}_{\rm T}$, where $\mathbf{a}_{\rm T}$ and  $\mathbf{a}_{\rm V}$  are the monopole and dipole coefficients in temperature and velocity units respectively. 
\end{enumerate}

We iterate the above pipeline to study the properties of the recovered bulk flows by (a) using different filters; (b) varying the amplitude of the input bulk flow velocity;  (c) using different cluster catalogs; (d) inputing different systematic components into the maps to determine their importance.

\subsection{Optical depth}

The calibration of conversion from dipole amplitude to flow velocity, through the calibration matrix $\mathbf{M}$ (section~\ref{ss:conv}), requires cluster position and optical depth. 
In this work, we assume optical depth is measured in the observed samples with negligible measurement errors and consider the scatter in the $Y-M$ relation for \planck\ and EASS clusters and $L-T$ relation for RASS clusters when producing our simulations (appendix~\ref{ss:scaling}). When dealing with clusters from observations, we have to derive the optical depth of the cluster sample to reconstruct the kinetic SZ signal. We briefly outline below how we can recover optical depth from SZ and X-ray clusters respectively.

For SZ observation, \planck, the optical depth of each cluster is directly obtained from their SZ flux, or equivalently the integrated compton Y-parameter, by $\tau=Y  (m_e c^2)/(k_B T_e)$ if the electron temperature $T_e$ is known. The temperature can be obtained from X-ray measurements through the $L-T$ scaling relations. For X-ray observations, RASS and EASS, we can calculate the optical depth from the electron density by $\tau=\sigma_{\rm T} \int n_{\rm e} dl$. We follow \osb\ to determine the cluster electron density (see equation 8 in \osb) using the Bremsstrahlung emission luminosity~\citep{Gronenschild1978} and the relation $T_{\rm X}=(2.76\pm0.08)L_{\rm X}^{0.33\pm0.01}$~\citep{White1997} to calculate the electron temperature of the cluster. For the RASS clusters, propagating the error in observed $L_{\rm X}$ to optical depth gives uncertainty of $\sim15\%$. The error in $L_{\rm X}$ for the EASS clusters is likely to be smaller since the background noise for eRosita is expected to be lower. The uncertainty in the optical depth for RASS is therefore an upper limit for EASS clusters. 

\section{Cluster Catalogs}
\label{s:cat}
We construct cluster catalogs for three representative all sky surveys: 
the combined ROSAT All-Sky \mbox{Survey} (hereafter RASS), the future \planck\ cluster survey and eRosita All-Sky Survey (hereafter EASS)\footnote{http://www.mpe.mpg.de/projects.html\#erosita}. 
Each survey has its own advantages: the RASS sample  already exists and has  been used for previous bulk flow measurements (e.g. KAKE and \osb\ ); the \planck\ cluster sample will contain about three--five times as many clusters, will extend to greater distances and,being an SZ survey, will be subject to different selection effects than RASS; the EASS cluster sample will be larger than both  the ROSAT and \planck\ samples and will extend to greater redshift $z\sim1$.
All three surveys provide isotropic samples outside regions at low galactic latitudes.
All surveys are flux-limited, and so at any redshift z only objects of mass $M>M_{\rm lim}(F_{\rm lim})$ are included in the catalog. A summary of the catalog properties is listed in Table~\ref{t:clu}.


All surveys are flux-limited, and so at any redshift z only objects of mass $M>M_{\rm lim}(F_{\rm lim})$ are included in the catalog. A summary of the catalog properties is listed in Table~\ref{t:clu}.

\begin{figure*}
  \begin{center}
    \leavevmode
       \includegraphics[width=100mm]{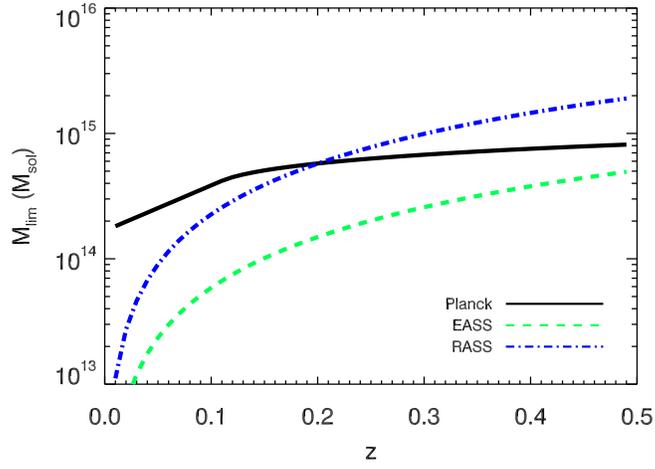}
       \caption{The minimum mass as a function of redshift for a cluster to be included in the RASS, \planck, and EASS catalogs. The cluster mass that we use in the catalogs is max[$10^{14}M_{\odot},M_{\rm lim}(z)$]. For the RASS sample, we fit equation~\ref{eq:xr} to the archived catalog and obtain the $M_{\rm lim}$ values shown in this plot.}
     \label{f:mlim}
  \end{center}
\end{figure*}

\begin{figure*}
  \begin{center}
    \leavevmode
       \includegraphics[width=100mm]{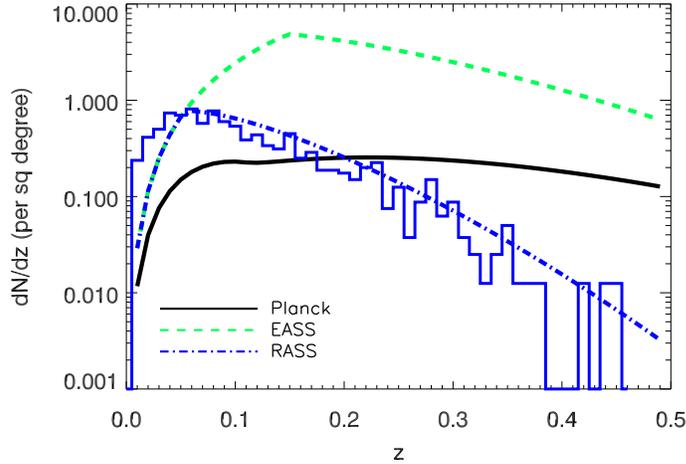} 
       \caption{The differential cluster redshift distribution per square degree for clusters in the \planck, EASS and RASS, with cluster mass $M>{\rm max}[10^{14} M_{\odot},M_{\rm lim}(z)]$. The $\Lambda$CDM cosmology is assumed and $\sigma_8=0.8$ is used for the number counts of the \planck\ and EASS clusters. The observed number count of the RASS clusters are overlaid with the solid lines for comparison with the fitted number counts (dash dot line). }
     \label{f:nofz}
  \end{center}
\end{figure*}

\subsection{eRosita All-Sky Survey}
The eRosita mission\footnote{mission definition document at http://www.mpe.mpg.de/erosita/MDD-6.pdf} is expected to be launched in 2012 and perform the first all-sky X-ray imaging survey in the X-ray energy range up to 5 keV with a limiting flux of $F_{\rm lim, 0.5-5\ keV}=1.6\times10^{-13}{\rm \ erg\ s^{-1}\ cm^{-2}}$. The EASS is expected to yield a catalog of a few tens of thousands of clusters out to redshift $z\approx1.3$. Multi-band optical surveys are planned to provide photometric and spectroscopic redshifts for the EASS clusters with masses above $3.5\times10^{14}M_\odot h^{-1}$~\citep{Cappelluti2010}. The second phase of the eRosita survey (the wide survey) will detect more clusters by using longer exposures than the first phase (the all sky survey). However, it will only cover about half of the sky resulting in a non-isotropic cluster sample and so we only use clusters expected from all sky survey.  We apply a galactic cut at $|b|\le7^\circ$ giving a sky fraction 0.727. The limiting mass of cluster to be included in the sample at redshift z is given by~\citep{Fedeli2009}:

\begin{equation}
\frac{M_{\rm lim,200}(z)}{10^{15}M_\odot h^{-1}} = \frac{1}{E(z)}\left[ 4\pi d_L^2(z)\frac{F_{\rm lim, 0.5-2 keV}/c_b}{1.097\times10^{45} \rm erg s^{-1}} \right] ^{1/1.554}
\label{eq:xr}
\end{equation}
\noindent where $E(z)\equiv H(z)/H_0$ is the normalized Hubble parameter, $d_L(z)$ is the luminosity distance and $c_b$ is the band correction factor which converts the bolometric flux to the eRosita energy $0.5-2$ keV. We estimate $c_b$ by assuming a Raymond-Smith~\citep{Raymond1977} plasma model with metalliticity of $0.4Z_\odot$, a cluster temperature of 4 keV, and a Galactic absorption column density of $n_{\rm H}=10^{21}\ {\rm cm^{-2}}$.  For consistency, we use the virial mass definition throughout this work and we convert $M_{200}$ to $M_{\rm v}$ (and write $M_{\rm v}$ as M hereafter) using the conversion fitting formula by~\citet{Hu2003}.

To create mock catalogs, we assign a mass and redshift to each cluster in the EASS sample and simulate the properties of the clusters using X-ray scaling relations (described in appendix~\ref{ss:scaling}).

\subsection{The \planck\ Catalog} 
\planck\ is imaging the whole sky with an unprecedented combination of sensitivity ($\Delta T/T\sim1\times10^{-5}$ at 143 GHz), angular resolution ($5'$ at 217 GHz), and frequency coverage ($30-857$ GHz). The SZ signal is expected to be measured from a few thousand galaxy clusters.  \planck\ will produce a cluster sample with median redshift $\sim0.3$ with significant fraction beyond redshift 1 (Planck Blue Book). As with EASS, we do not simulate clusters with galactic latitude $|b|\le7^\circ$ to minimize the Galactic signal. The SZ observable is the integrated Comptonization parameter $Y=\int y\ d\Omega_{\rm cluster}$ (appendix~\ref{ss:scaling}) with expected values of $Y_{\rm 200}>10^{-3}{\rm arcmin^2}$~\citep{Schafer2007}, where $Y_{\rm 200}$ is the integrated comptonization parameter within $r_{200}$.~\citet{Fedeli2009} provide a fitting formula for the limiting mass as a function of redshift based on simulations by~\citet{Schafer2007}:
 
\begin{equation}
\frac{M_{\rm lim,200}(z)}{10^{15} M_\odot h^{-1}}=
\begin{Bmatrix}
e^{-1.200 + 1.469{\rm \tan^{-1}}[(z-0.11)]^{0.440}]} \ {\rm if\ z\ge0.11} \\
e^{-1.924 + 8.333z} \ {\rm if\ z\le 0.11}
\end{Bmatrix}
\label{eq:pla}
\end{equation}

\subsection{ROSAT All-Sky cluster survey}
The RASS sample, consisting of clusters from the REFLEX~\citep{Bohringer2004}, BCS~\citep{Ebeling2000a}, eBCS~\citep{Ebeling2000b}, CIZA~\citep{Ebeling2002,Kocevski2007}, and MACS~\citep{Ebeling2001} catalogs comprises a total of 827 clusters. All of the clusters have spectroscopic redshifts with $z<0.5$. We simulate the SZ signal of the clusters in this catalog from the observed X-ray properties via scaling relations, while the redshifts and positions on the sky are taken from the catalog. 

Figure~\ref{f:mlim} shows the limiting mass for clusters to be included in the RASS catalog. We fit equation~\ref{eq:xr} to the 827 RASS clusters and find an effective flux limit of $F_{\rm lim}=1.2\times10^{-12} {\rm ergs^{-1}\ cm^{-2}}$. In Figure~\ref{f:nofz} we show the cluster number counts $dN/dz/d\Omega$, for our cluster catalogs.

\begin{deluxetable}{lcccc}
  \tablecaption{Parameters of cluster catalogs}
  \tablewidth{0pc}
  \tablehead{
   \colhead{Catalog} &\colhead{RASS}&\colhead{\planck} &\colhead{EASS} \\
     \colhead{units} &\colhead{$f_{\rm 0.5-2 keV} $}& \colhead{$Y_{200} $} &\colhead{$f_{\rm 0.5-2 keV} $} \\
     \colhead{} & \colhead{$ergs^{-1}cm^{-2}$} &\colhead{${\rm arcmin^2}$} & \colhead{$ergs^{-1}cm^{-2}$} 
  }
  \startdata
  Flux limit & $1.2\times10^{-12}$&$10^{-3}$ & $1.6\times10^{-14}$ \\
  $z_{\rm median}^{\ast}$& 0.091 &0.246 &0.211 \\
  $\left \langle N_{\rm cl} \right \rangle$ & 827 & 2700 & 33000 \\
    $\left \langle \tau \right \rangle$ &  $(7.1\pm2.2)\times10^{-3}$ & $(8.5\pm1.7)\times10^{-3}$ & $(6.1\pm1.5)\times10^{-3}$
  \enddata 
  \tablecomments{$^\ast$ The median redshift is computed from the simulated cluster sample with a cut-off redshift z=0.5 for all three cluster samples.}
    \label{t:clu}
\end{deluxetable}

\section{Sky Simulations}
\label{s:szmap}
We create full-sky SZ maps using a semi-analytical approach. We follow the method presented in  in~\citet{Delabrouille02} and~\citet{Waizmann09} that used the halo model and cluster gas properties. The simulations include SZ emission, primary CMB anisotropy and instrument noise, while diffuse Galactic foregrounds are excluded.

Clusters are simulated with properties expected from the surveys described in section~\ref{s:cat} with number densities given by the halo mass function $N(z,M)$. Each simulated cluster is then assigned a mass and redshift and the gas properties are then derived. We leave the details of the number counts and the cluster models to appendix~\ref{s:sim} and focus on the preparation of the SZ maps here. 

In the nonrelativistic limit the distortion to the CMB temperature of the kinetic SZ effect is given by equation~\ref{eq:ksz} but with the peculiar velocity replaced by $\vec{v}=\vec{V}_{\rm bulk} + \vec{v}_{\rm peculiar}$. We give an overall bulk velocity $\vec{V}_{\rm bulk}$ to the whole cluster sample which is a free parameter of the simulation. Besides a bulk motion, we also give each cluster a random peculiar velocity $\vec{v}_{\rm peculiar}$ drawn from a Gaussian distribution with variance given by equation~\ref{eq:sigv} with $R=8 \mpch $.

Maps of CMB anisotropies are generated from the angular power spectrum using the CMBfast code~\citep{Seljak1996} for a flat $\Lambda$CDM cosmology.  The CMB and SZ maps are combined and then smoothed with a Gaussian beam with sizes given in Table~\ref{t:table_planck_channel}. Finally, the instrument noise is added with the noise variance and the beam size of the \planck\ channels are listed in Table~\ref{t:table_planck_channel}. Synthesized maps are generated in the HEALPix pixelization scheme~\citep{Gorski2005}, with resolution of Nside=1024. 

\section{Filtering: Reconstruction of the Kinetic SZ signal }
\label{s:filter}
The kinetic SZ signal from a cluster is embedded in the CMB  and instrument noise,  and tSZ emission is also present at the same location.  We therefore filter the maps to increase the signal-to-noise of our cluster dipole measurement. 
We use two types of linear multifrequency filters to reconstruct the cluster kinetic SZ signal. The filters are constructed with the aim of minimizing the CMB and noise variance in the map. We leave the derivation of the filter shapes to the appendix~\ref{a:filter}. The first filter is a matched filter (hereafter MF;~\citealt{Haehnelt1996,Herranz2002}) that is optimized to detect the kSZ signal. The second filter (hereafter UF) is subject to the additional constraint of removing the tSZ signal at the cluster location. The UF was first proposed by~\citet{Herranz2005} for use on flat patches of the sky and later~\citet{Schafer2006b} extended the scheme for use on full sky maps. Both filters use all of the frequency channels and take into account the statistical correlation of the CMB and instrument noise between different frequencies.

\begin{figure}
  \begin{center}
    \leavevmode
       \includegraphics[width=90mm]{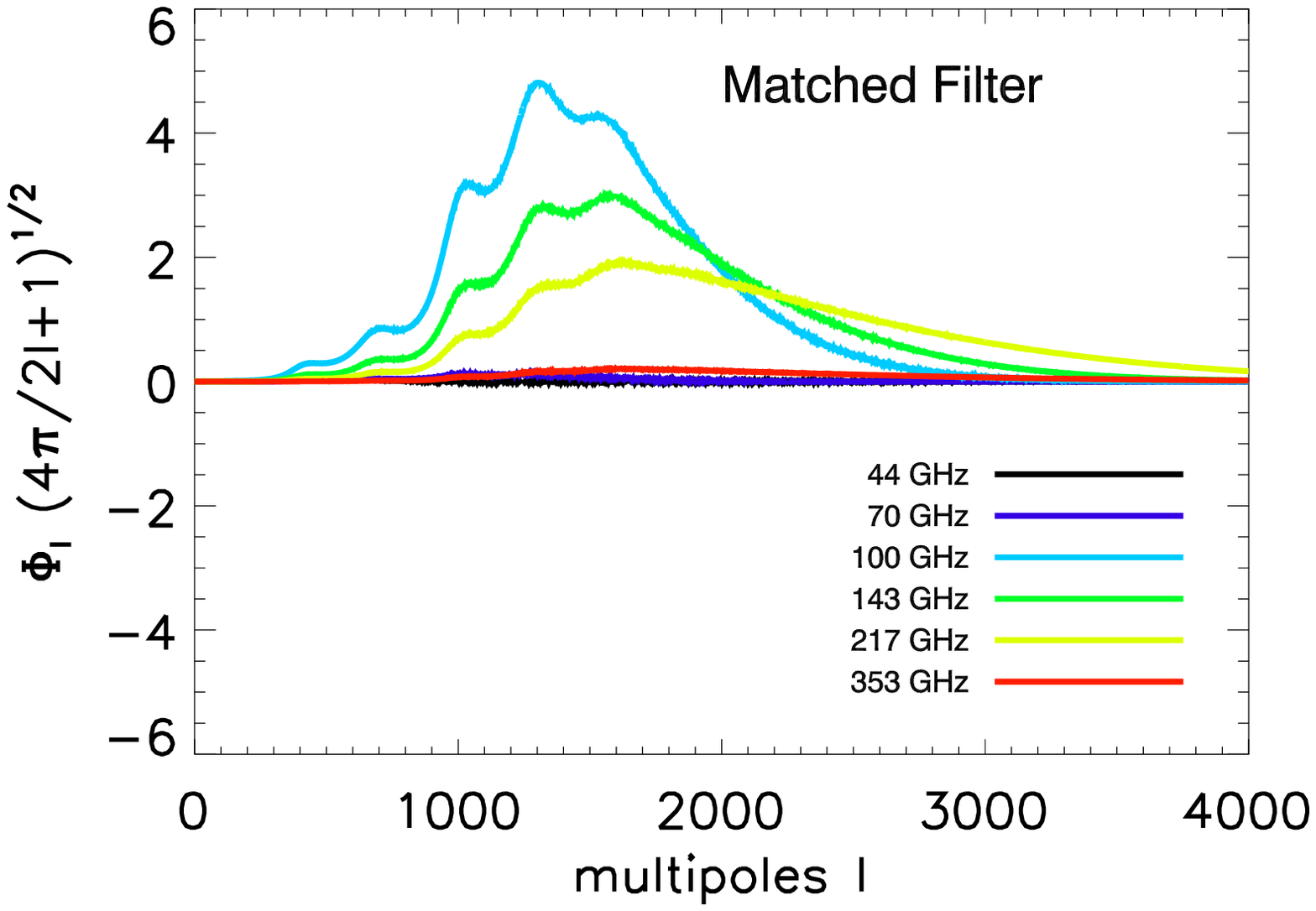} 
       \includegraphics[width=90mm]{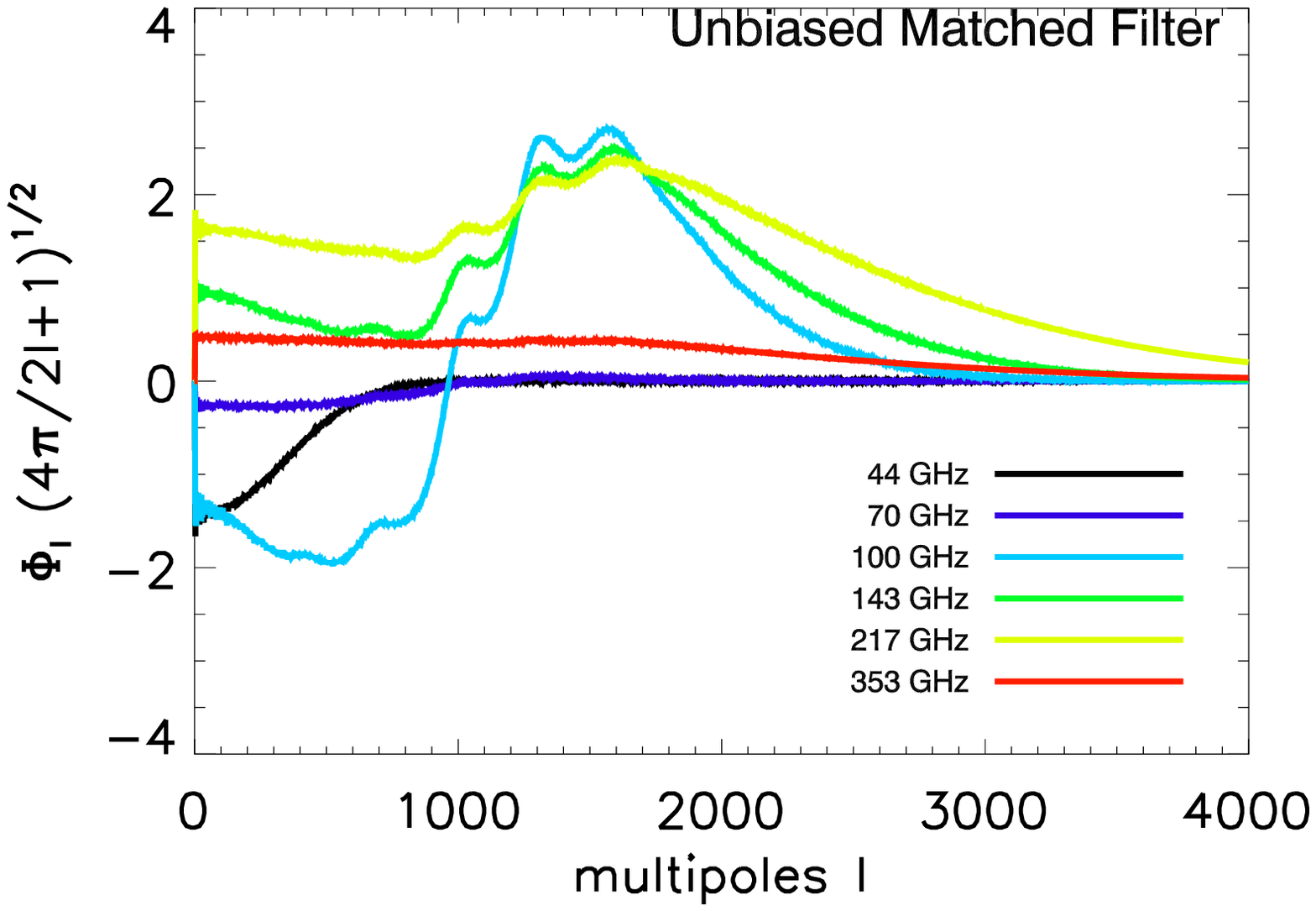} 
       \caption{(Upper) Multi-frequency matched filter signed to detect the kSZ signal. (Lower) Unbiased multi-frequency matched filters designed to remove the tSZ components . Both filters are constructed using the \planck\ beam profiles as the cluster radial profiles.  }
     \label{f:filters}
  \end{center}
\end{figure}

The filter kernels are:

\begin{equation}
 {\rm matched\ filter}: \ \mathbf{\Phi}_l^{\rm MF}=\frac{\mathbf{C_l}^{-1}}{\gamma}\mathbf{B_l} 
\label{eq:mf} 
\end{equation}
\begin{equation}
 {\rm unbiased\ filter}: \ \mathbf{\Phi}_l^{\rm UF}=\frac{\mathbf{C_l^{-1}}}{\Delta}(\alpha \mathbf{B_l} - \beta \mathbf{F_l}) 
\label{eq:kszumm} 
\end{equation}

\noindent where $\alpha=\sum^{l_{max}}_{l=0}{\mathbf{F_l^{\rm T}C_l^{-1}F_l}}$, $\beta=\sum^{l_{max}}_{l=0}{\mathbf{B_l^{\rm T}C_l^{-1}F_l}}$, $\gamma=\sum^{l_{max}}_{l=0}{\mathbf{B_l^{\rm T}C_l^{-1}\mathbf{B_l}}}$, and $\Delta=\alpha \gamma - \beta^2$. 

We assume the clusters are point sources in the \planck\ maps and so $\bold{B_l}$ are the spherical harmonic coefficients of the \planck\ beam function, $\bold{F_l}= f(\nu)\bold{B_l}$  where $f(\nu)$ gives the tSZ frequency dependence, $\bold{C_l}=\bold{C_l^{\rm CMB}} + \bold{C_l^{\rm noise}}$ is a matrix at every multipole giving the sum of the cross power spectra of the CMB and instrument noise between frequency channels. The filters are normalized such that the filtered field gives the amplitude of the kinetic SZ signal at the central pixel of each cluster. 
Given the spatial resolution of \planck\, we compute the filter kernels out to multipole $l_{\rm max}=3000$. Figure~\ref{f:filters} shows the MF and UF for the 6 \planck\ frequency channels we use, using the \planck\ beam FWHM and pixel noise variance.

\subsection{Matched filter (MF)}
The MF is less complicated than the UF since most of the features are only seen at the cluster scales ($1000\leq l \leq 2500$) while the smallest and largest scales are suppressed. As with the UF, the channels $\nu=100$, $\nu=143$, $\nu=217$ GHz are given the largest weights. The matched filter is more efficient than the UF at removing the CMB and instrument noise components because it forces the scales at which the noise and CMB dominant to disappear. However, since the MF is not designed to remove the thermal SZ signal, the amplitudes of the filter kernels are all positive. 

\subsection{Unbiased Kinetic SZ filter (UF)}
The UF is constructed to give an unbiased estimate of the kinetic SZ signal at the clusters' locations. At large and intermediate scales ($l\leq1000$) the channels below $\nu=100$ GHz are subtracted from those above. The CMB fluctuations which dominate the signal at these scales are therefore suppressed. At cluster scales ($1000\leq l\leq2500$) all of the filter kernels have positive amplitudes and the $100$, $143$, $217$ and $353$ GHz channels are given the largest weight. At 100 GHz and 143 GHz the tSZ signals are negative, approximately zero at 217 GHz and positive at 353 GHz. The channels are weighted so that the tSZ signal is zero. Although the tSZ signal is smaller at 100 GHz than the other three channels, the pixel noise level is lower ($\sigma_{\rm N}\approx6\mu K$) and so the 100 GHz channel also has substantial weight. At smaller scales ($l\geq2500$) the signal is dominated by instrument noise and so all channels are suppressed by the filter.

\section{Analysis Pipeline}
\label{s:dipole}

\subsection{Weighted Least Square Fitting of Dipole}
We fit the real spherical harmonic coefficients of the monopole, $a_0$, and dipole terms, $a_{1x}$, $a_{1y}$ and $a_{1z}$, to the filtered maps using a weighted least square fit which is based on the Healpix IDL procedure {\it remove\_dipole}~\citep{Gorski2005}:
 
\begin{equation}
\mathbf{a_{\rm T}=(X^{T}WX)^{-1}X^{T}Wu}
\label{eq:leastsquare}
\end{equation}

\noindent where $\mathbf{a_{\rm T}}$ is a vector of best fit monopole and dipole coefficients, $\mathbf{u}$ is the filtered map (e.g.equation~\ref{eq:ff}), $\mathbf{W}$ is a matrix with diagonal elements equal to the weight given to each pixel of the map, and $\mathbf{X}$ is a matrix giving the contribution of the fitting function to each pixel. We give the central cluster pixel of the map a weight $W_i=1/\sigma_{N,i}^2$, where $\sigma_{N,i}$ is the i--th pixel noise variance, all other pixels are given zero weight. The noise variances are calculated from 100 filtered CMB and noise realizations.  We also tried weighing each pixel by an estimate of the signal to noise, i.e. $W_i=\tau_i/\sigma_{N,i}^2$, but we find a larger systematic bias due to tSZ contamination in the recovered bulk flow than when using the former weighting scheme. We therefore use a weighting scheme that involves only the pixel noise.

We fit the dipole only at central cluster pixels because our filters are optimized to reconstruct the source amplitude if the source were centered at that pixel.

\subsection{Translating from $\mu K$ to km/s}
\label{ss:conv}
We construct a $4\times4$ matrix $\mathbf{a}_{\rm V}=\mathbf{Ma}_{\rm T}$ to convert the dipole from temperature units ($\mu K$) to velocity units (km/s), in which the matrix elements having units of km/s/K. This involves the use of simulated kinetic SZ maps that can account for the attenuation of the optical depth by the beam convolution and the filtering process. In principle one can perform this unit conversion in observations, e.g. \planck\ temperature sky maps, using calibrated SZ simulations based on estimation of the optical depth of each clusters. This is the case in \osb\ that they used the simulated kSZ maps of the RASS clusters to calibrate \wmap\ data. On the other hand, in this work we assume the optical depth, hence the kSZ signal, of each cluster is known and calculate the matrix $\mathbf{M}$ from our sky simulations. Note that the matrix elements depend implicitly on the average optical depth of the cluster sample, therefore it is specific to individual cluster experiment and we construct it separately for the three cluster surveys. We now describe how we create this matrix.

We generate four different sets of kinetic SZ signal only realizations with a given bulk flow amplitude: one with a monopole velocity and the other three with a dipole velocities in the x, y, z directions.  We choose an amplitude of of 10,000 km/s, which is large enough such that the recovered signal is not confused by uncertainty of optical depth and random peculiar velocity. Since the kinetic SZ signal of each cluster is proportional to the optical depth, the elements of the matrix $\mathbf{M}$ depend implicitly on the average optical depth of the cluster sample. Each set of maps are passed through the pipeline to obtain the $a_{1m}$. The four $a_{1m}$ from each set of realizations are the elements of each row of $\mathbf{M}^{-1}$. By repeating this procedure for the four kinetic SZ signal only realizations, we construct the 16 elements of the calibration matrix. We check that the off-diagonal elements of $\mathbf{M}$ are at most $1\%$ of the diagonal ones for all of the cluster samples considered. 

We perform this exercise 20 times and assign the average value from these 20 realizations to the matrix elements. For each of the RASS, \planck, and EASS sample, we calculate one average $\mathbf{M}$ and apply it to the filtered temperature maps. Due to the intrinsic scatter of the $Y-M$ relation introduced in the simulations, the average optical depth of any simulated cluster sample might deviate from the expected value from the characteristics associated with that catalog, this may lead to calibration error (section~\ref{ss:pec}).  
   
\section{Error estimation}
\label{s:error}
The error budget to the dipole coefficients contains several terms:

\begin{equation}
\sigma_{\mathbf{a_{\rm V}}}^2=\sigma_{\rm CMB+noise}^2 +  \sigma_{\rm kSZ}^2  +  \sigma_{\rm tSZ}^2  + \sigma_{\rm PS}^2
\label{eq:errors}
\end{equation}

\noindent where $\sigma_{\rm CMB+noise}$ is the residual from the CMB and instrument noise, $\sigma_{\rm kSZ}^2$ is the error in the kinetic SZ signal due to the intrinsic scatter of the $Y-M$ relation and the random component of the galaxy cluster peculiar velocity that is not part of the bulk flow, $\sigma_{\rm tSZ}^2$ is the thermal SZ residual, and $\sigma_{\rm PS}^2$ is the contamination from radio or infrared point sources associated with clusters. In principle, the last three may be correlated, while we are assuming here that correlations are negligible.

The errors on the three dipole coefficients $a_{1m}$ are correlated and the uncertainty in the best fit value can be described by a covariance matrix $\mathbf{N}$. We compute the covariance matrix of each type of error in equation~\ref{eq:errors} by passing simulations of the components through our pipeline and performing the dipole fit on them. The scatter $\sigma$ in dipole coefficients then provides an estimate of the noise correlations between the dipole directions, i.e. $\mathbf{N}=\left \langle \mathbf{a_{\rm V}a_{\rm V}^{T}} \right \rangle$. Then,

\begin{equation}
\chi^2=\mathbf{(a_{\rm V,rec}-a_{\rm V,in})^{\rm T}N_{\rm tot}^{-1}(a_{\rm V,rec}-a_{\rm V,in})}
\label{eq:chi}
\end{equation}

\noindent where $\mathbf{a_{\rm V,rec}}$ is the best fit dipole coefficients of the recovered velocity and $\mathbf{a_{\rm V,in}}$ is the input velocity dipole coefficients. The confidence level of a given $\mathbf{a_{\rm V}}$ can be calculated from the $\chi^2$ probability distribution with 3 degrees of freedom.  

\begin{figure}
  \begin{center}
    \leavevmode
       \includegraphics[width=90mm]{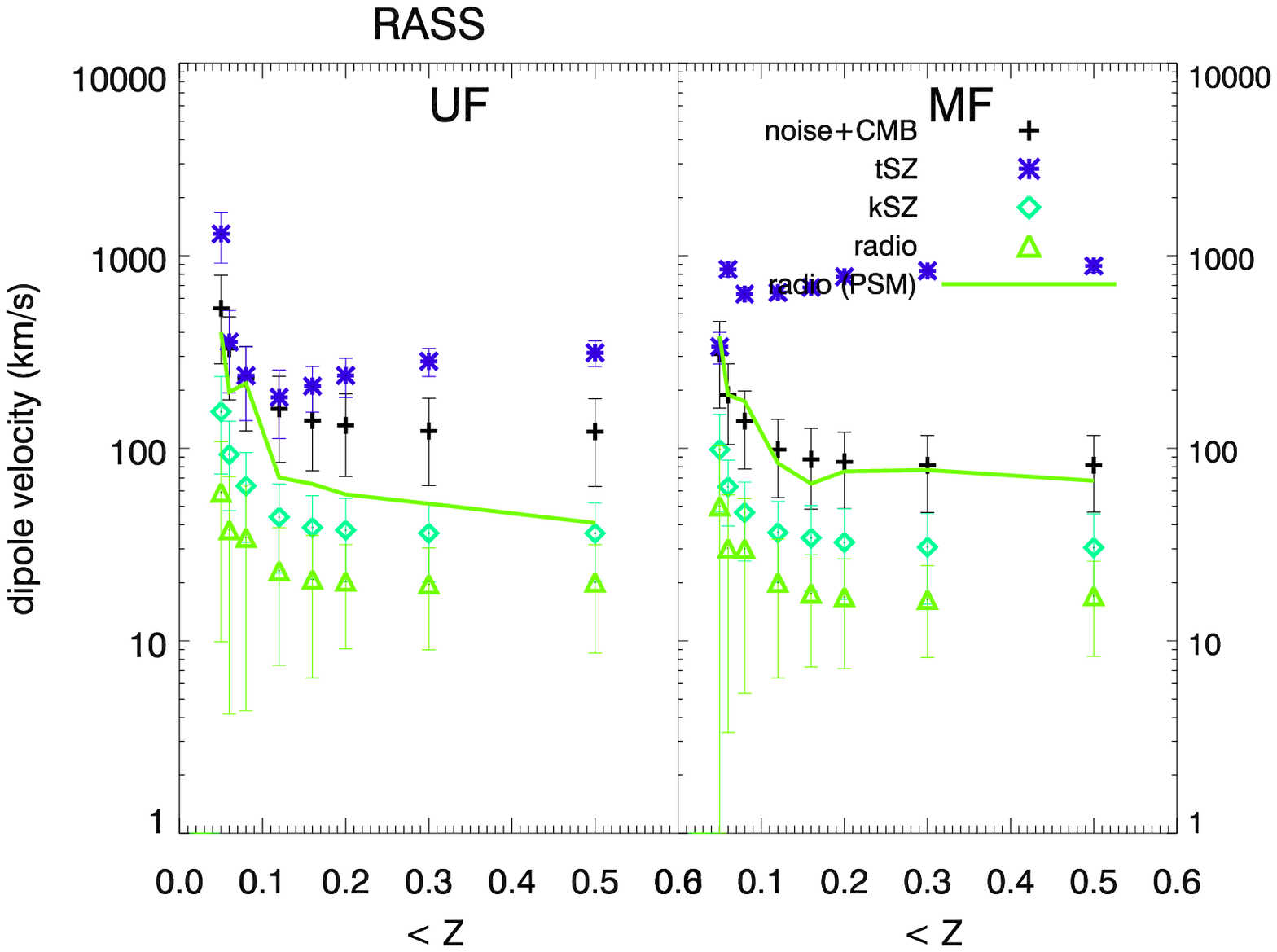} 
       \includegraphics[width=90mm]{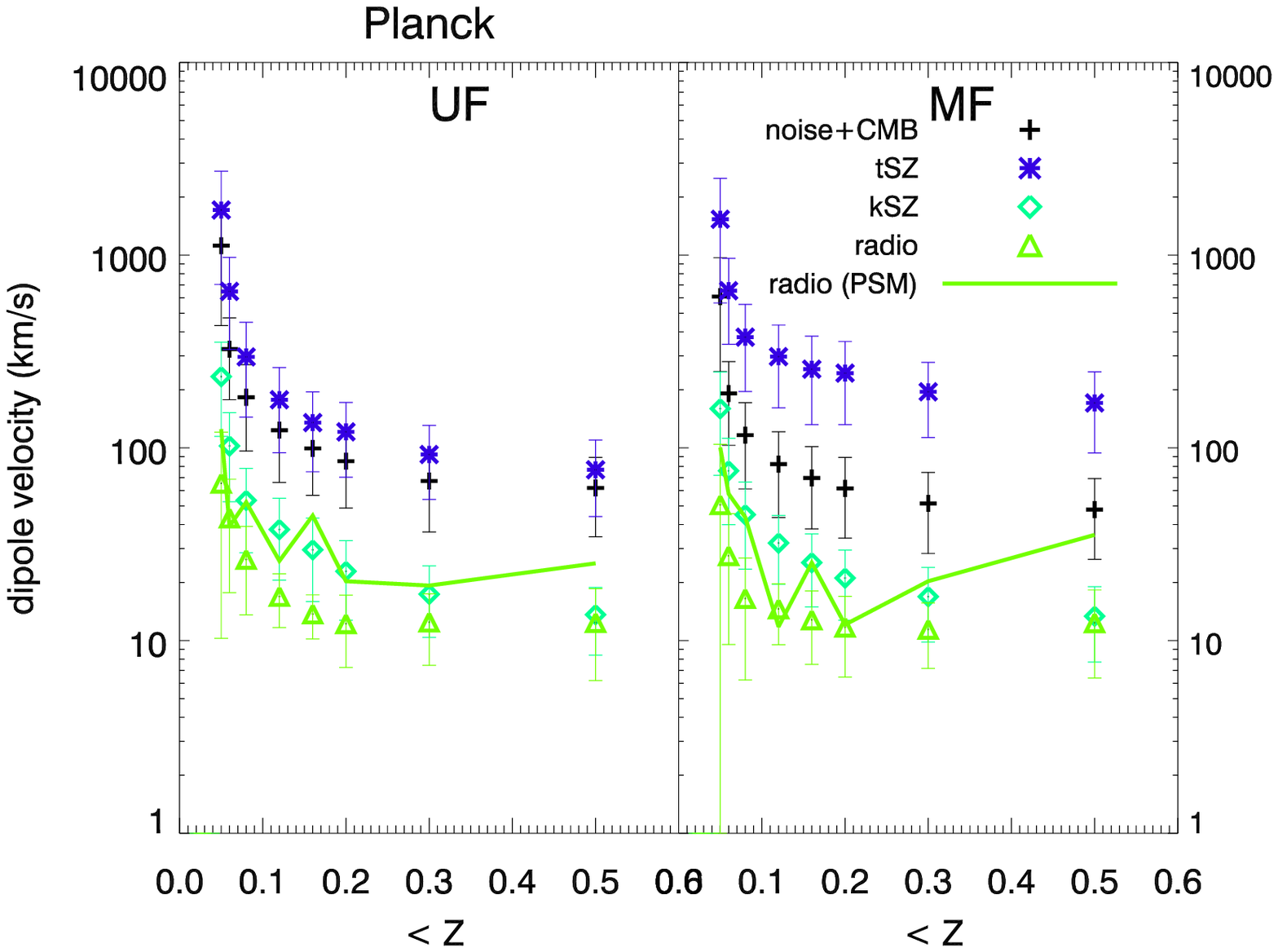} 
        \includegraphics[width=90mm]{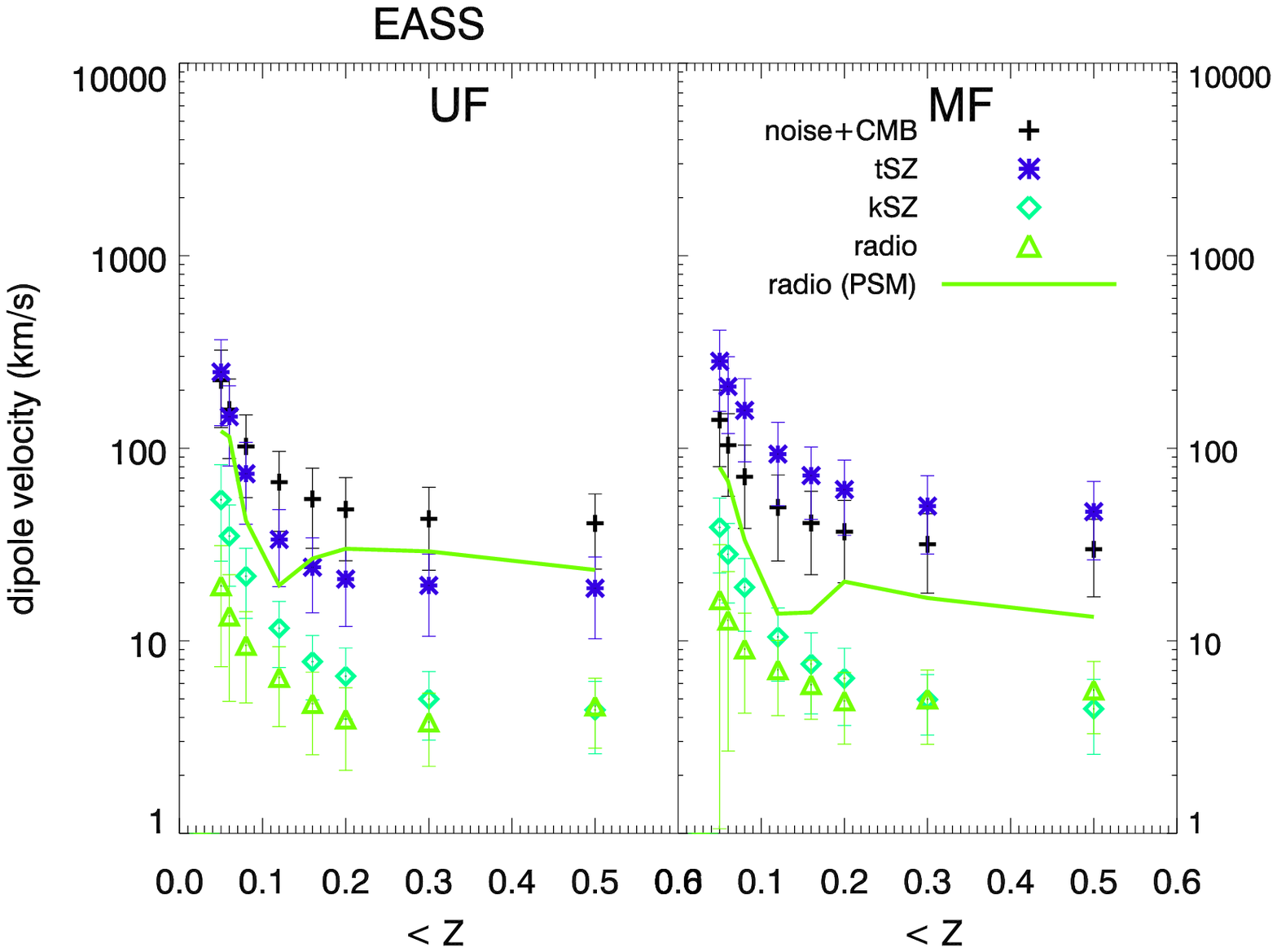} 
       \caption{The systematic error contributions to the bulk flow measurement using the RASS (top), \planck\ (middle), and EASS (bottom) cluster samples:  instrument noise and CMB (cross), thermal SZ (asterisk), uncertainty in the kinetic SZ signal due to scatter in both the optical depth and cluster peculiar velocity (diamond), and the contamination from radio point sources (triangle for our simulation and solid line for PSM simulation). The data points are the average values from 100 realizations and the error bars are the standard deviations of the one-sided distribution. }  
     \label{f:errors}
  \end{center}
\end{figure}

Figure~\ref{f:errors} shows the contribution from each systematic effect to the dipole velocity. We discuss these results in the following subsections. 

\subsection{CMB and instrument noise}
\label{ss:noisecmb}
The CMB signal is not completely removed by our filters because there is CMB emission on cluster scales with the same frequency dependence as the kSZ signal. We evaluate the error from the CMB and instrument noise by measuring the bulk flow of   300 filtered CMB plus noise maps. 
As a further check on this procedure we also calculate the dipole on a single CMB plus noise realization and repeat the procedure with 100 realizations of the cluster sample, each having a different spatial distribution. The results from the two procedures are similar with errors that are both gaussian with similar variance.

As expected the UF errors are larger than  the MF ones by about $30\%$ for the three cluster samples. We find that the error is largest for the RASS sample and smallest for the EASS sample, since the EASS sample contains more clusters.

\subsection{Thermal SZ signal }
\label{ss:thermalbias}
We expect the thermal SZ signal to be entirely removed when using the unbiased matched filter (UF). However, maps filtered with the MF suffer from a systematic bias by the unfiltered thermal SZ signal that contaminates the kinetic SZ signal. In addition to the effect of filtering, the intrinsic scatter of the cluster y-parameter would also introduce further incomplete removal of the thermal signal and resulting in scatters around the mean recovered velocities. To evaluate the level of thermal SZ bias and uncertainty, we generate simulated maps containing only tSZ signal.

We find that in thermal SZ signal only maps the average systematic bias ($\pm1\sigma$ uncertainty as represented by the error bars in Figure~\ref{f:errors}) to the cluster dipole are $v=313\pm48$ km/s for RASS, $v=77\pm33$ km/s for \planck, and $v=19\pm8$ km/s for EASS when filtered by the UF, and $v=885\pm66$ km/s for RASS, $v=171\pm77$ km/s for \planck, and $v=47\pm20$ km/s for EASS when filtered by the MF. The thermal bias is most serious for MF filtered RASS clusters while the $1\sigma$ uncertainty is the largest for MF filtered \planck\ clusters. This is due to the fact that the average optical depth is the largest  for \planck\ clusters and hence the largest intrinsic scatter (and uncertainty). 

We find a significant monopole: $v_{a_0}=$ 730 (RASS), 1440 (\planck), and 870 (EASS) km/s  when filtered by the UF, and $v_{a_0}=$ 3500 (RASS), 3890 (\planck), 2180 (EASS) km/s when filtered by the MF. The UF is more effective at removing the thermal SZ signal than the MF, by a factor of $\sim 5$. Though the monopole velocities are large compared to the thermal dipole, they have no effect on the bulk flow measurement.

\subsection{Uncertainty in optical depth and random cluster peculiar velocity}
\label{ss:pec}
We find that the cluster random peculiar velocity not part of the bulk flow gives a negligible dipole with an average magnitude of $v<50$ km/s for all three cluster samples and is independent of the type of filter used. The error due to uncertainty in the optical depth depends on the amplitude of the peculiar velocity since $\sigma_{\rm kSZ}\propto v \sigma_{\tau}$. We find that $\sigma_{\tau}\approx15\%$ for the \planck\ cluster sample and $\sigma_{\tau}<15\%$ (O10) for the RASS and EASS clusters. The uncertainty for \planck\ clusters is relatively larger  because it contains more abundant massive clusters and hence propagate larger errors to the optical depth for the same scatter in the cluster Y-parameter. As mentioned earlier, this uncertainty introduces a calibration error when translating from $\mu$K to km/s because of the difference of the average optical depth and the calibration matrix $\mathbf{M}$. However, since we choose a large calibration velocity of 10,000 km/s, the calibration error is only about $5\%$ and negligible when compared to the error due to CMB and noise. 

\subsection{Extragalactic Point Sources}
\label{ss:radio}
At small scales extragalactic point sources are significant contaminants of CMB maps. They give a Poisson noise contribution to the measured angular power spectrum and a non-Gaussian signature in the maps~\citep{Colombo2010}. Radio point sources have a falling spectrum with $\alpha\sim0.7$ where $S\propto\nu^{-\alpha}$, whereas infrared point sources have a rising spectrum with typical spectral indices $\alpha\sim3$. Thus infrared sources tend to be the more significant contaminants at higher frequencies where the SZ effect is large. Recent measurements  such as the South Pole Telescope (SPT) detected 47 IR sources at S/N$>4.5$~\citep{Vieira2010}. Nevertheless, the infrared point source signal is less well characterized than the radio point source signal and reliable models are not available to estimate the number counts within a galaxy cluster. Furthermore, the 353 GHz channel, which suffers the most from IR point source emission is given less weight than the other channels at cluster scales. We therefore do not consider the contribution from infrared sources in this work.  Radio sources are an important contaminant of SZ measurements as they fill in the SZ temperature decrement if the sources are within the cluster being observed, leading to an underestimation of the SZ signal. We expect a negligible contribution since the cluster signal is strongest at frequencies where the radio point source signal is small.

To estimate the level of radio contamination to the cluster kSZ dipole we simulate maps of the radio point source emission. To do this we use the radio luminosity function at 1.4 GHz of~\citet{Lin2007} extrapolated to higher frequencies using, 

\begin{equation}
\frac{dn(\log P_{\nu})}{d\log P_{\nu}}=\int{ \frac{dn(\log P_{1.4})}{d\log P_{1.4}}f(\alpha) d\alpha}
\label{eq:rlf}
\end{equation}
\noindent where $P_{\nu}$ is the radio luminosity in units of ${\rm WHz^{-1}}$at frequency $\nu$, $\alpha$ is the spectral index, and $f(\alpha)$ is the probability distribution of the spectral indices at 1.4 GHz, which is taken to be a Gaussian distribution with mean $\bar{\alpha}=0.51$ and rms $\sigma_{\alpha}=0.54$. $f(\alpha)$ is taken to be $f(\alpha+0.5)$ for $\nu>90$ GHz to account for possible steepening of $\alpha$ at high frequencies. The RLF at 1.4 GHz is given by the fitting formula:

\begin{equation}
\log \left ( \frac{dn}{d\log P_{1.4}} \right )=u-\sqrt{b^2+ \left (\frac{\log P_{1.4}-x}{w} \right )^2}-1.5\log P_{1.4}
\label{eq:rlf14}
\end{equation}
\noindent where $u=37.97$, $b=2.40$, $x=25.80$, and $w=0.78$~\citep{Lin2007}.  Here $n$ is the radio source number density within $r_{200}$. We integrate equation~\ref{eq:rlf} to obtain the expected number of radio sources, $N_{\rm radio}$, in a cluster and assign fluxes to the $N_{\rm radio}$ sources using the RLF. Since bright sources will be masked in the \planck\ maps we also mask sources with $S>S^{\rm lim}_\nu$, where $S^{\rm lim}_\nu$ is the upper limit of the radio flux at frequency $\nu$ given by~\citet[Table 2]{Lopez2006} leaving a radio map with sources having $S<500mJy$.

We find no significant cluster dipole from radio point sources in the RASS, \planck, or EASS cluster samples. In the redshift shell $z=0-0.5$ we find signals of $v<10$ km/s for EASS and $v\sim20$ km/s for both \planck\ and RASS clusters when filtered with either the UF or MF. We can further verify the result for the RASS sample by using more realistic radio point source simulations that considers information from sources observed by NVSS, and simulations from the Planck Sky Model\footnote{\url{http://www.apc.univ-paris7.fr/APC_CS/Recherche/Adamis/PSM/psky-en.php}} (PSM). As before, we mask bright radio sources in the PSM simulation. The result from the PSM simulation is shown in Figure~\ref{f:errors}. The radio dipoles from the PSM are consistently larger than our radio simulations at all redshift shells, but are of a similar order of magnitude with $v\sim50$ km/s. 

\section{Results}
\label{s:results}
We generate sets of simulations containing CMB, noise, thermal SZ signal, and kinetic SZ signal with 68 bulk flow velocities logarithmically spaced between 100 and 10,000 km/s. In this section we describe how well we recover the input bulk flow.

\subsection{Precision and detection threshold of Bulk Flow Measurement}
\label{ss:precision}
\begin{figure*}
  \begin{center}
    \begin{tabular}{cc}
        \includegraphics[width=85mm]{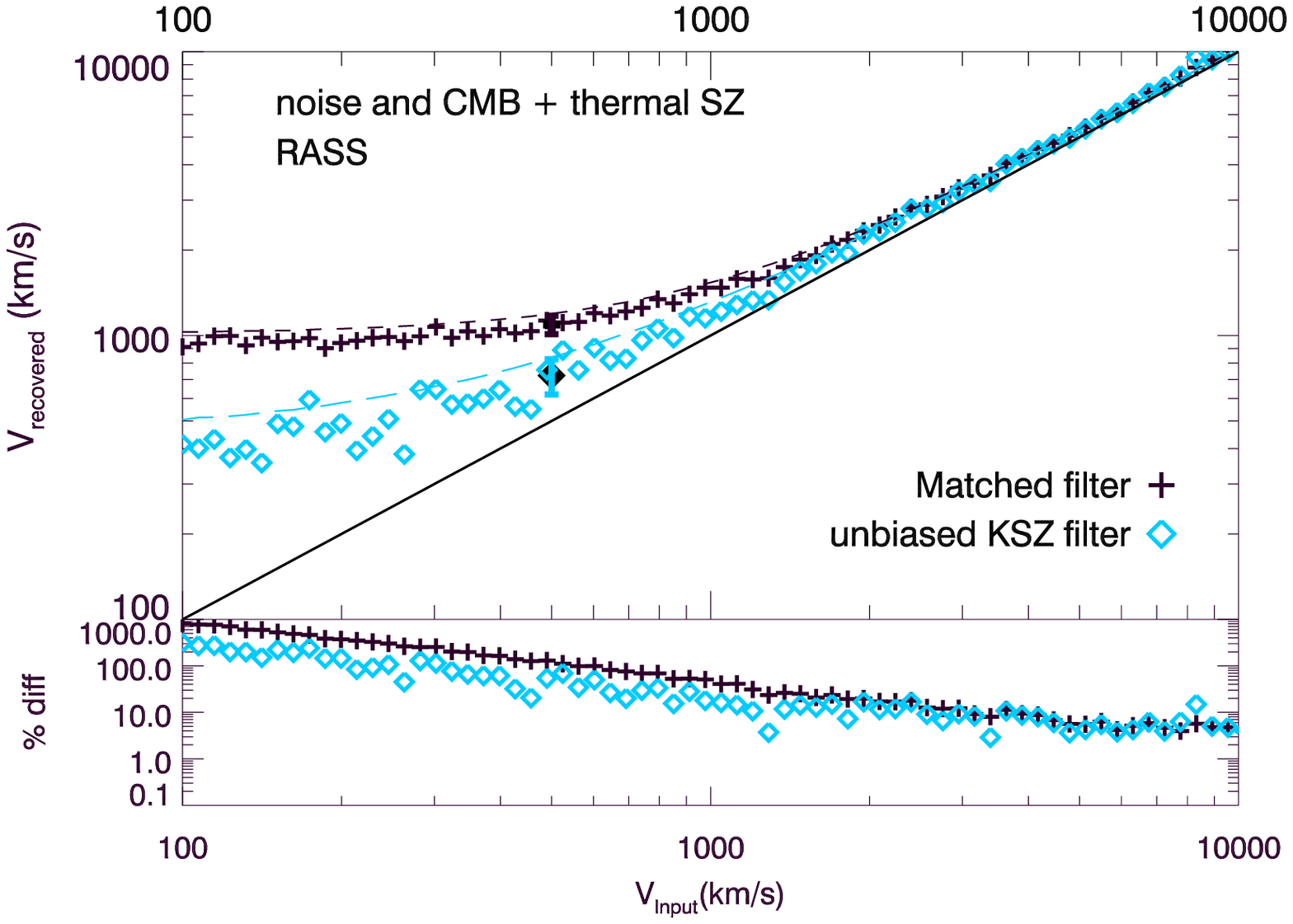}  &
         \includegraphics[width=85mm]{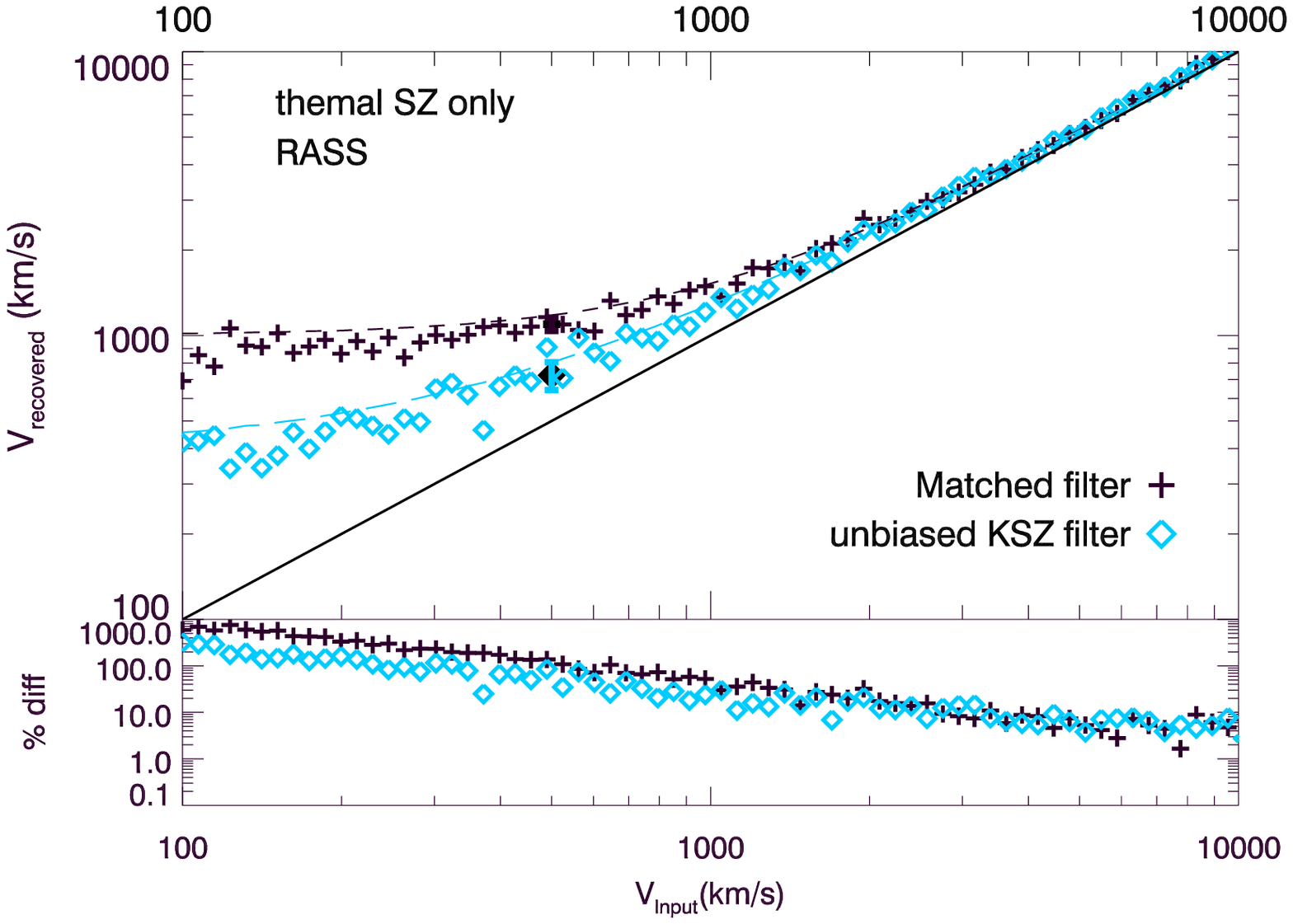}  \\ 
         \includegraphics[width=85mm]{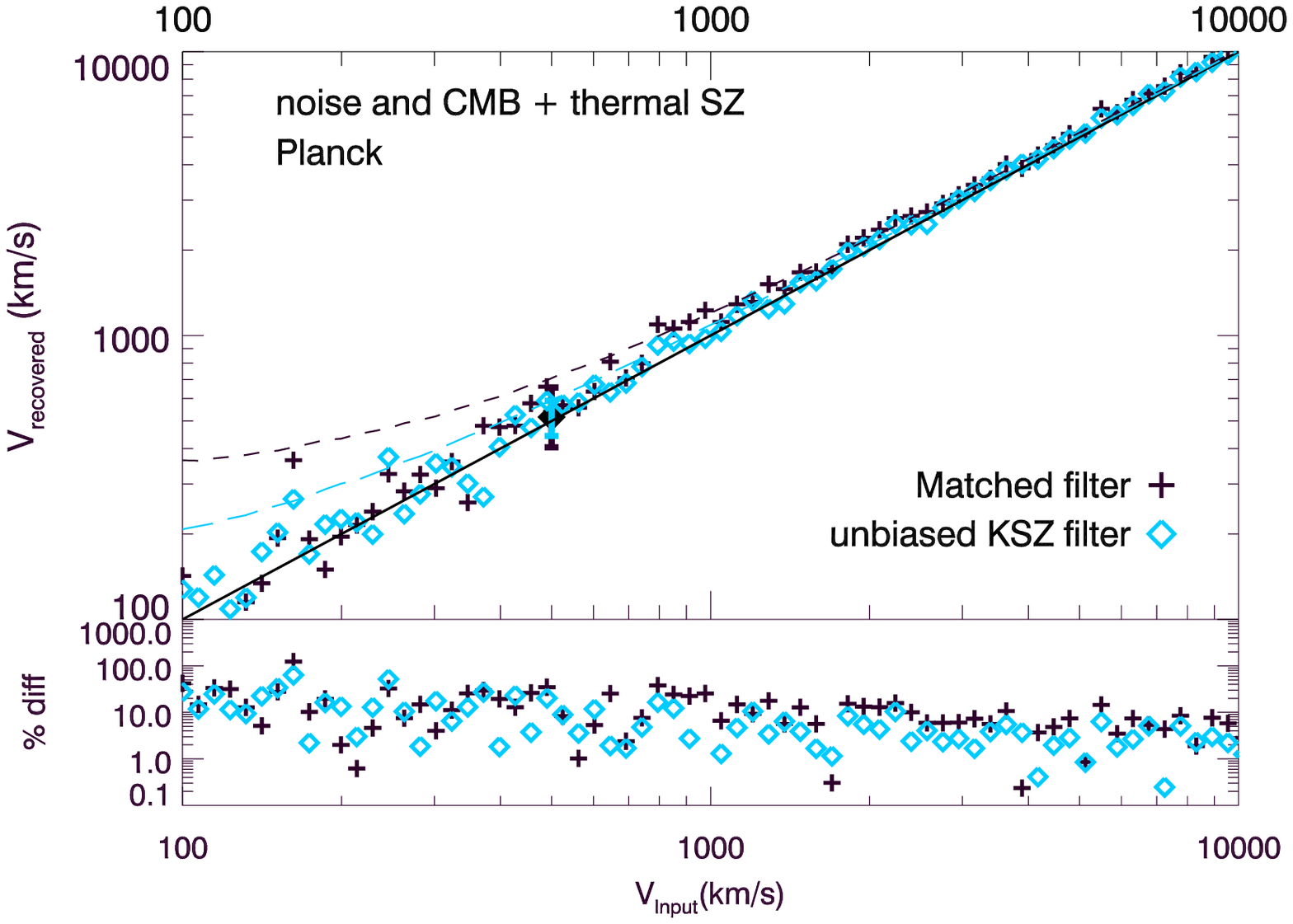}  &
         \includegraphics[width=85mm]{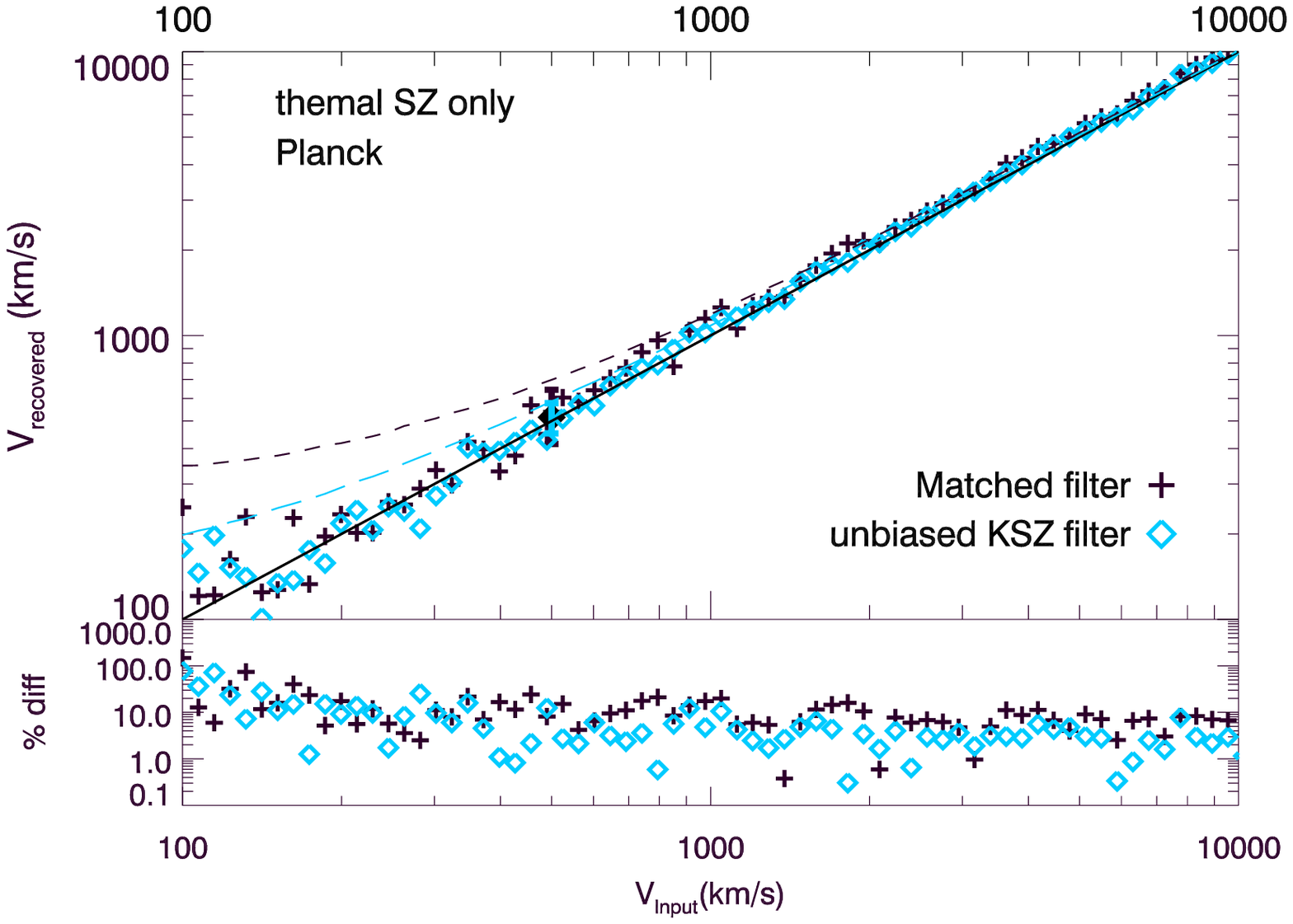}  \\     
        \includegraphics[width=85mm]{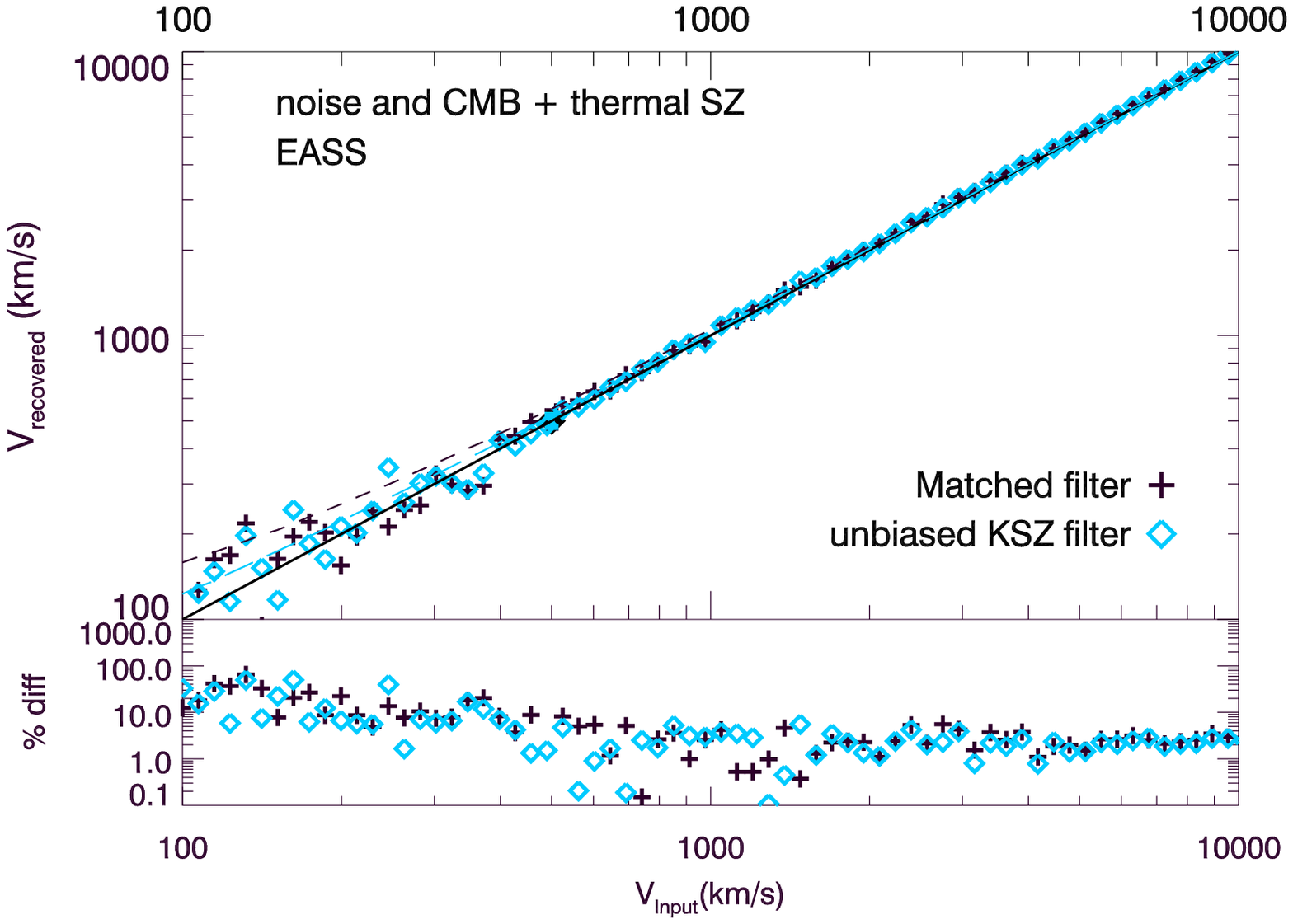}  &
         \includegraphics[width=85mm]{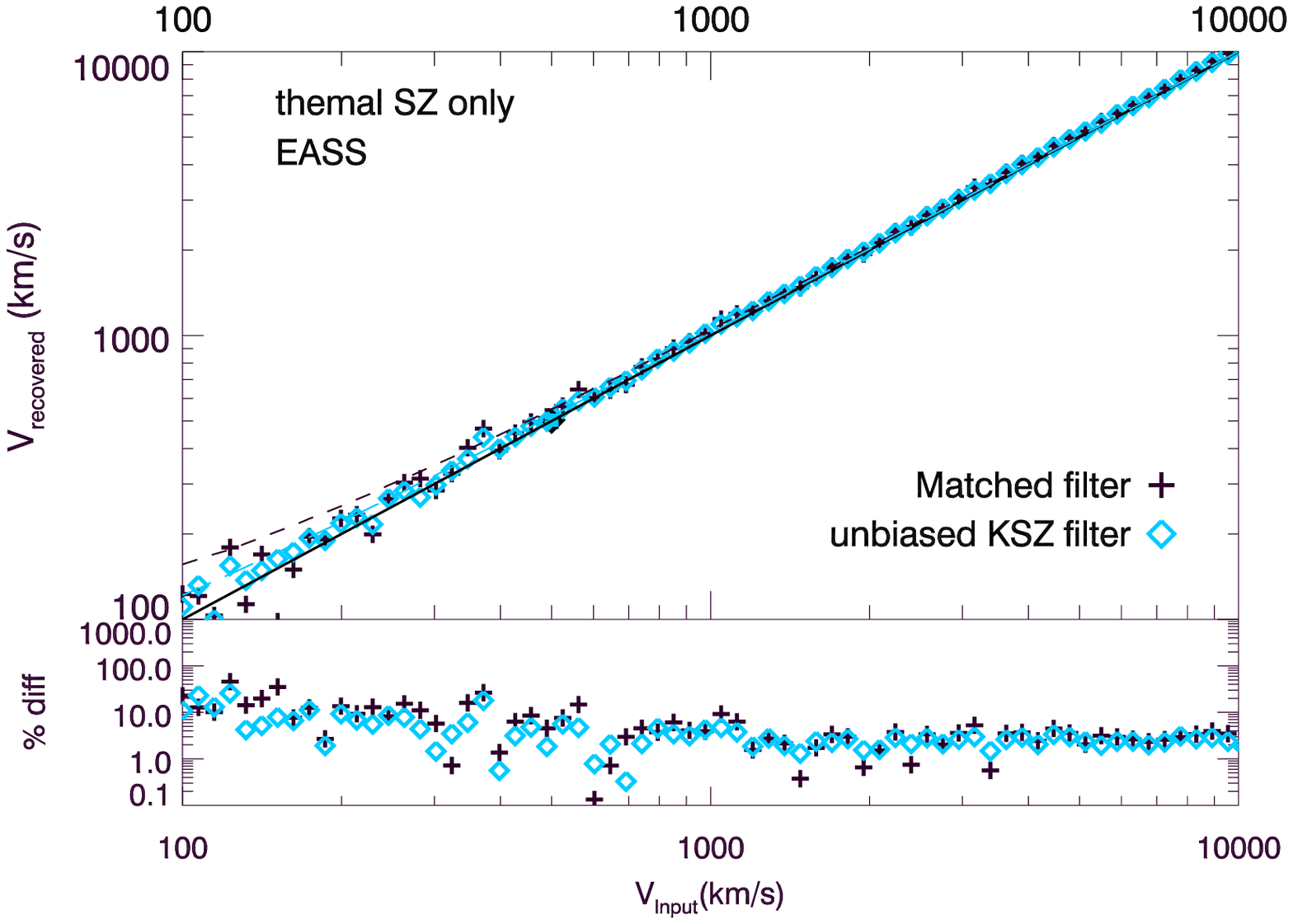}  \\      
     \end{tabular}
       \caption{Recovered bulk flow velocity for clusters $z\le0.5$ ( $\mathit{Top}$: RASS, $\mathit{Middle}$: \planck, $\mathit{Bottom}$: EASS) from simulated maps with input bulk flow velocities logarithmically spaced between 100 and 10,000 km/s, filtered with the MF (black cross) and UF (blue diamond). The solid line represents perfect recovery. The dash lines quote the 95\% confidence interval of the input bulk flow values. All results are based on maps containing kinetic and thermal SZ signals, but simulated with different systematic components. Plots labelled Noise and CMB are from maps containing instrument noise and CMB; plots labelled thermal SZ are from results containing thermal SZ emission and hence estimate the level of thermal SZ bias. }
     \label{f:vpplanck}
  \end{center}
\end{figure*}

\begin{figure*}
  \begin{center}
    \leavevmode
       \includegraphics[width=100mm]{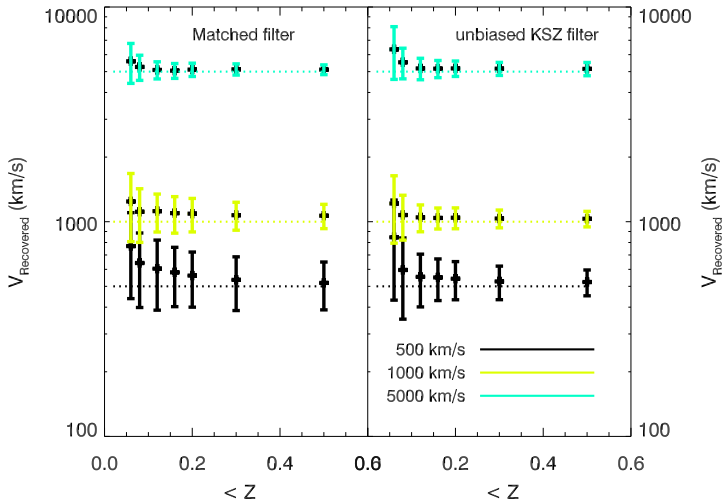} 
       \caption{Recovered bulk flow velocity for three input bulk flows $v= $500 (black), 1000 (yellow), and 5000 (cyan) km/s as a function of redshift for the \planck\ cluster sample. The recovered bulk flows are measured from simulations that contain all systematic errors we consider.}
     \label{f:vpredshift}
  \end{center}
\end{figure*}

Figure~\ref{f:vpplanck} shows the recovered bulk flow values $V_{\rm recovered}$ versus the input values $V_{\rm input}$ for all clusters with redshift $z\le0.5$. The dashed lines in the figures are the $95\%$ upper limits to the recovered bulk flow velocities and are computed by generating 10,000 Gaussian distributed dipole coefficients at each of the x, y, and z directions having mean equal to the input bulk flow velocity and thermal bias, and variance equal to the noise levels described in section~\ref{s:error}, i.e. $a_{1m}^{\rm sim}=(a_{1m}^{\rm in}+a_{1m}^{\rm bias})\pm a_{1m}^{\rm noise}$. The resulting magnitude of the dipole velocity follows the $\chi^2$ distribution, the magnitude and angle error (section~\ref{ss:angle}) at $95\%$ are then computed.

We determine the precision of the velocity measurement by computing the deviation of the values $V_{\rm recovered}$ from their corresponding $V_{\rm input}$ with a parameter defined to quantify the deviations as follows:

\begin{equation}
\sigma_{\log V}= \sqrt{ \frac{ \sum_{i=0}^N(\log V_{\rm recovered, i} - \log V_{\rm input, i})^2 }{N-2} }
\label{eq:dispersionvv}
\end{equation}

\noindent where the summation runs over the range of $V_{\rm input}$ being considered and $N=68$. This parameter allows us to compare the level of dispersion among the results from the various experiments and surveys. We provide the values for each cluster sample in Table~\ref{t:summary}. 

Taking a representative velocity of $V_{\rm input}$ =500 km/s we find that the recovered velocities are overestimated by $120\%$ with a scatter of 16\% when evaluated using the RASS clusters using the MF ($V_{\rm recovered}= 1085\pm82$ km/s), but perfectly recovered with only a scatter of 5\% when evaluated on the EASS clusters using the UF ($V_{\rm recovered}= 500\pm24$). The large overestimation in MF filtered RASS clusters is due to tSZ contaminations which has the largest effect for the RASS clusters. We find that the recovered velocity has the largest uncertainty for MF filtered \planck\ clusters, with a scatter of 25\%, and is consistent with the error analysis that the uncertainty due to intrinsic scatter in Y-parameter is largest for \planck\ clusters. With the sensitivities we find, measurements using either the \planck\ or EASS cluster sample will be able to detect a bulk flow as large as that claimed by KAKE. If the bulk flow is consistent with the $\Lambda$CDM prediction of $v=30$ km/s for redshift shell extending to $z=0.5$, our analysis pipeline would recover a 95\% upper limit to the velocity of, when the UF is used, $v=470$ km/s for RASS cluster, $v=160$ km/s for \planck\ cluster sample and $v=60$ km/s for EASS cluster sample.

Figure~\ref{f:vpredshift} shows the recovered signal in different redshift shells at three input bulk flow velocities: $V_{\rm input}=$ 500, 1000, and 5000 km/s, estimated from 100 realizations. For illustration purpose, only the result from the \planck\ cluster sample is shown. At these velocities, the recovered velocities are shown to be consistent with the input values within $1\sigma$ at redshifts $z=0.1$. At lower redshifts, the recovered velocities are noise dominated. At this redshift limit, the \planck\ cluster catalog will contain approximately 400 clusters. If a bulk flow of amplitude $\approx500$ km/s is present our results suggest that a sample of about 400 clusters would allow \planck\ observations to constrain the bulk flow. 

\subsection{Recovering the Bulk Flow Direction} \label{ss:angle}
\begin{figure*}
  \begin{center}
    \begin{tabular}{cc}
        \includegraphics[width=80mm]{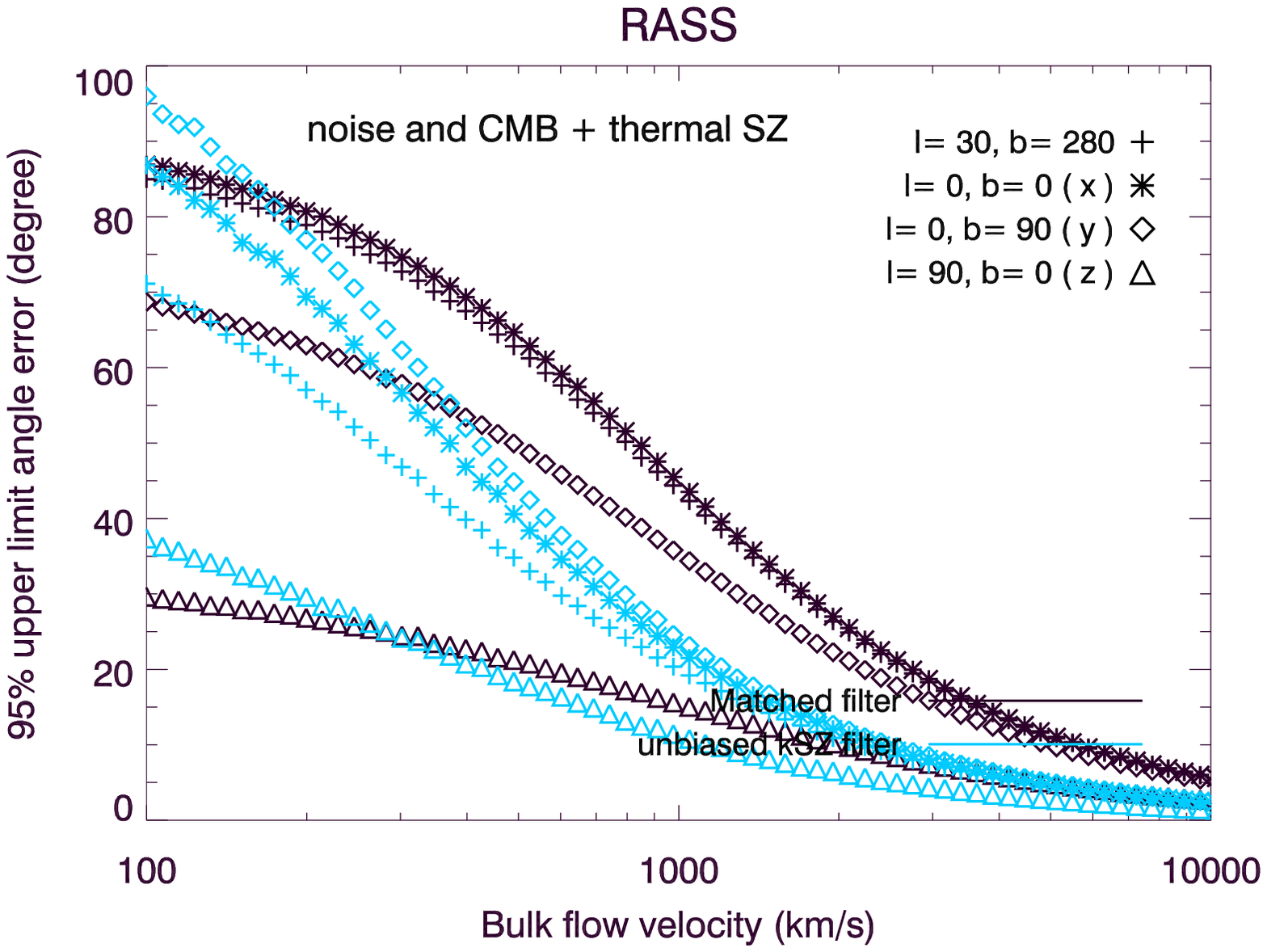}  &
        \includegraphics[width=80mm]{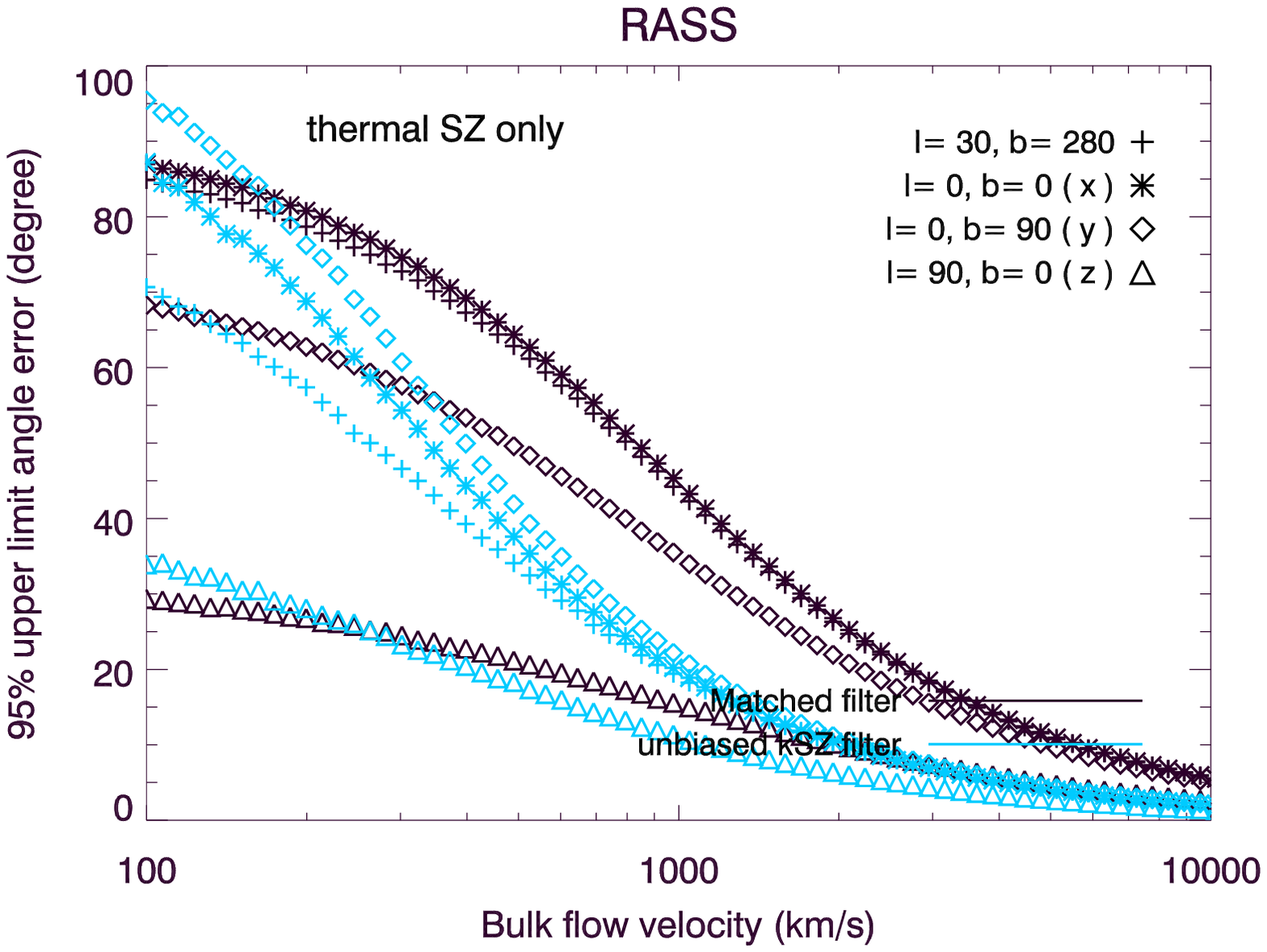}  \\
        \includegraphics[width=80mm]{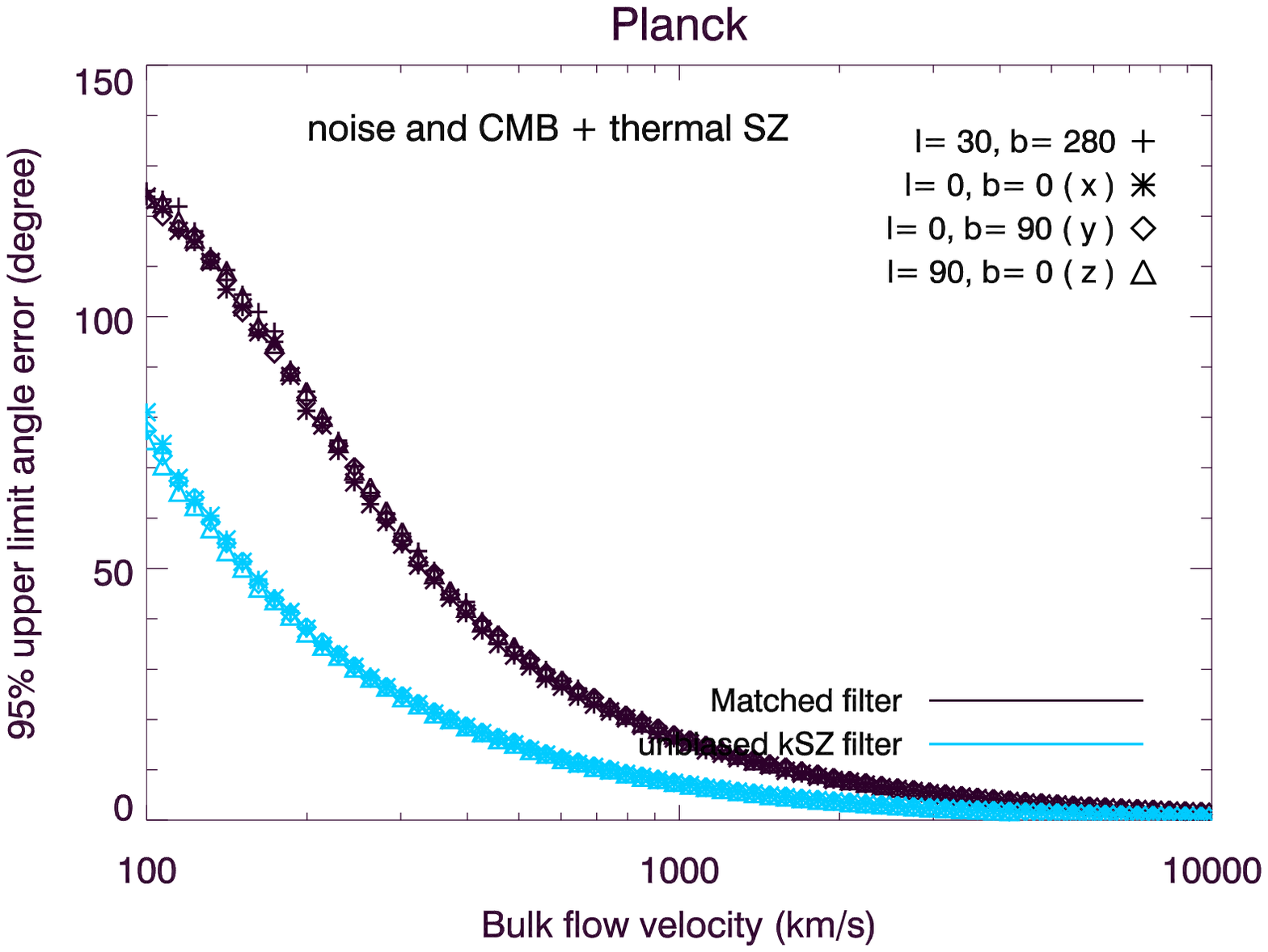}  &
        \includegraphics[width=80mm]{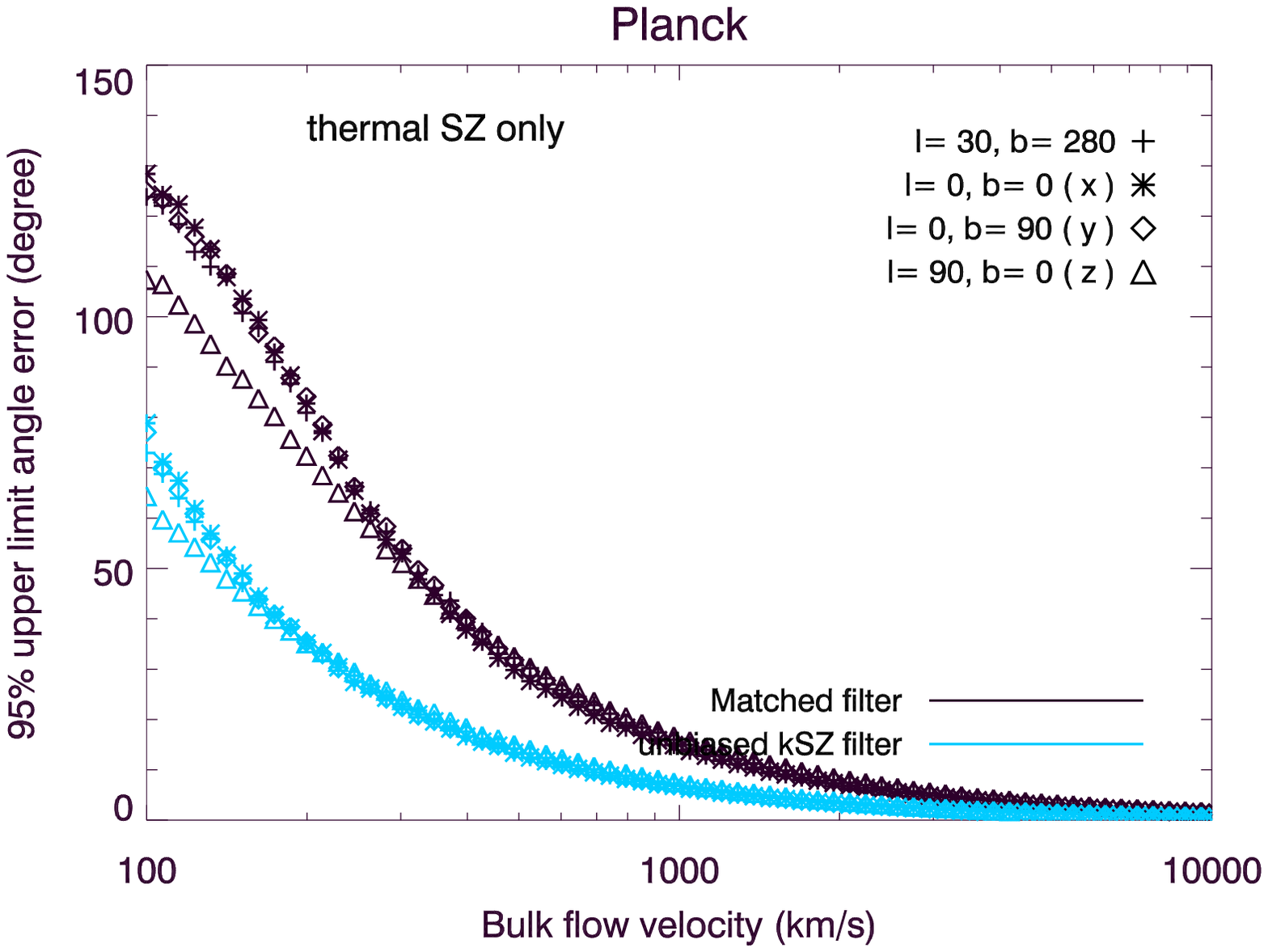}  \\
        \includegraphics[width=80mm]{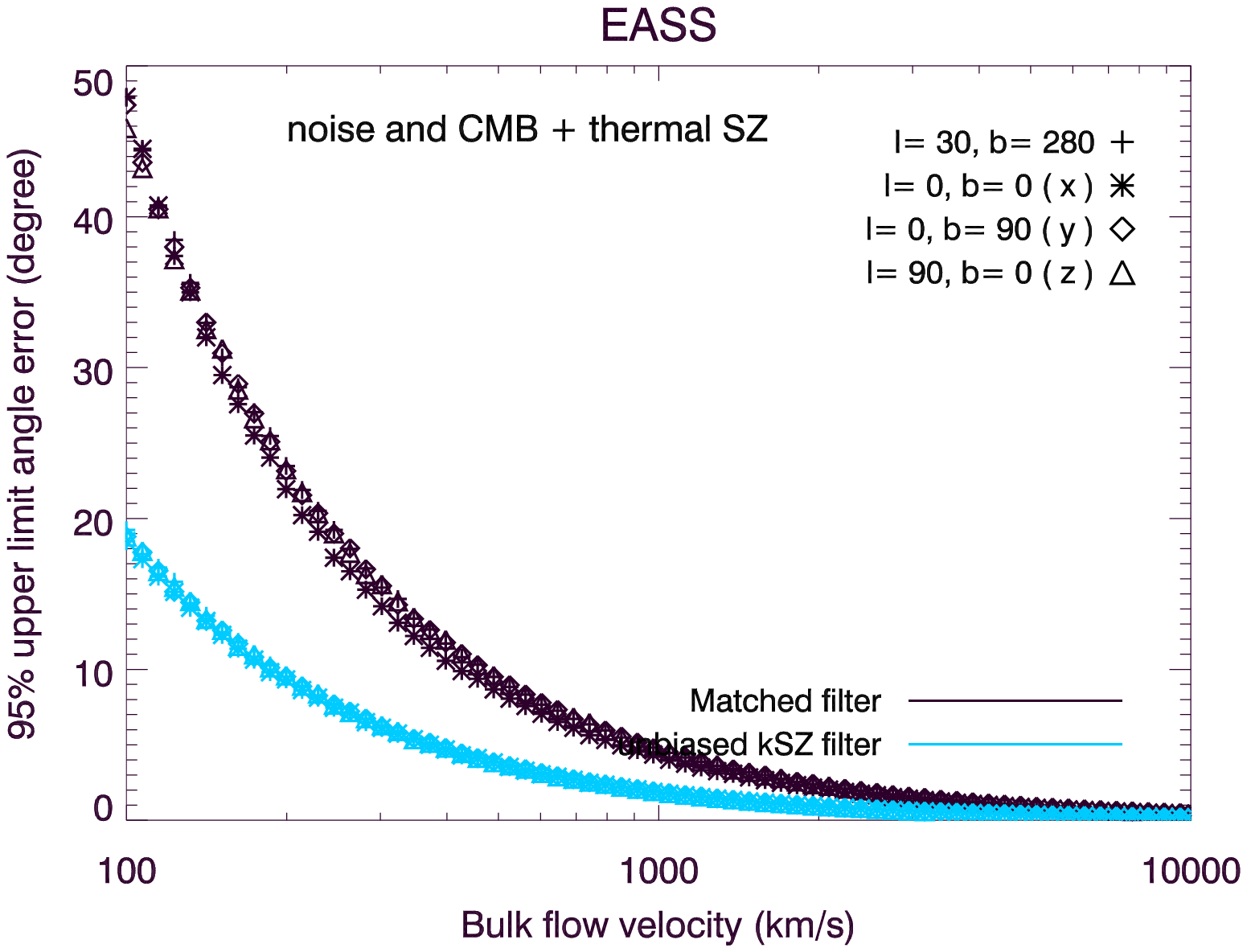}  &
        \includegraphics[width=80mm]{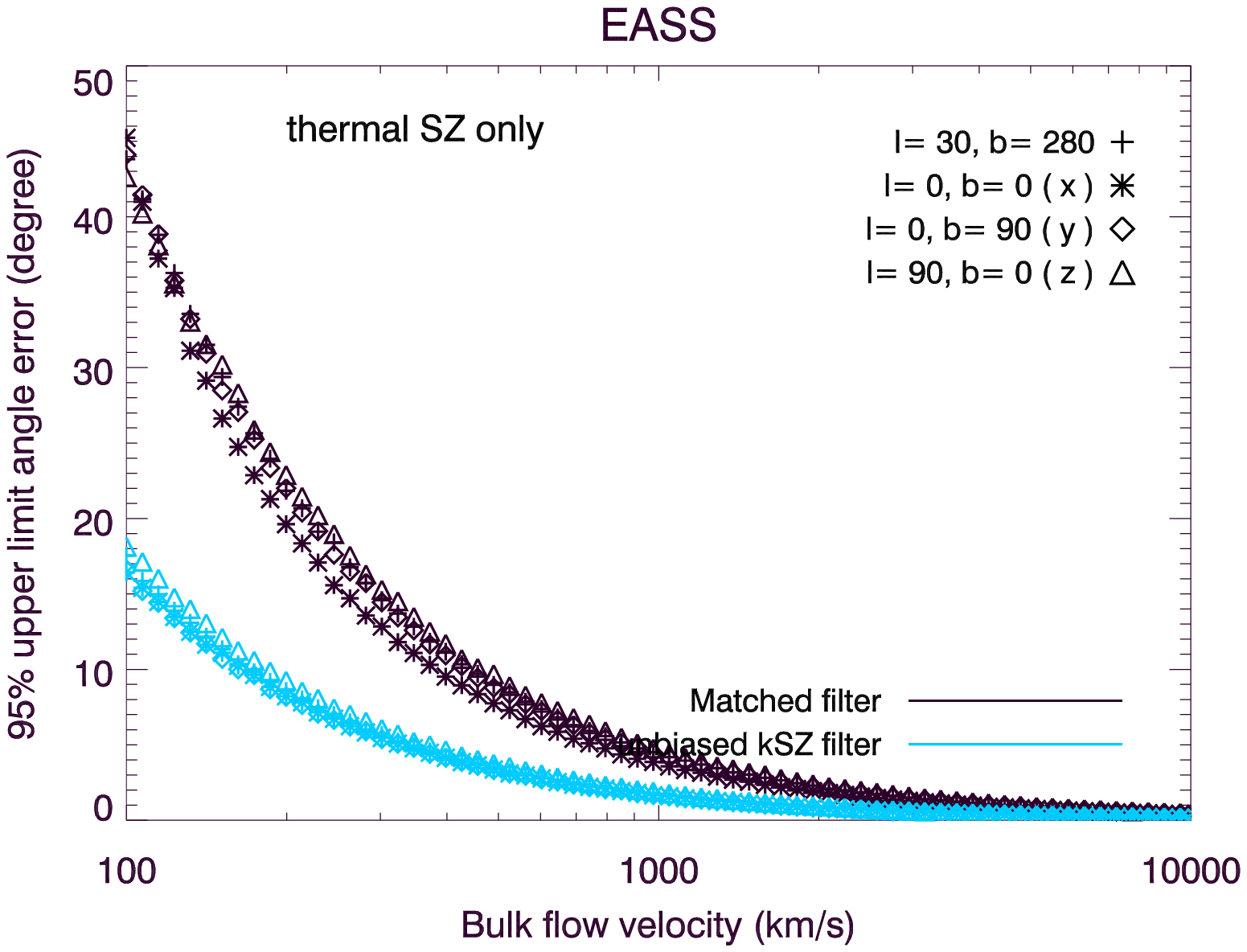}  \\       
     \end{tabular}
       \caption{The 95\% upper limit to the error in the angle measurements for all the three cluster surveys. The $l=30^{\circ}$, $b=280^{\circ}$ data points correspond to the direction of the input bulk flow that the simulations in Figure~\ref{f:vpplanck} are based on.}
      \label{f:chi95_angle}
  \end{center}
\end{figure*}

The direction of any bulk flow is not predicted by the $\Lambda$CDM model. We can estimate the error in the direction of a detected dipole for each of the three surveys. For $V_{\rm input}=500$ km/s towards $l=280^{\circ}$ and $b=30^{\circ}$ the average values of the dipole components, ($\left \langle V_x \right \rangle$,  $\left \langle V_y \right \rangle$, $\left \langle V_z \right \rangle$) from 100 simulations, are listed in Table~\ref{t:summary}. The direction is recovered with deviation $<20\%$ for the \planck\ and EASS cluster samples (except for $\left \langle V_x \right \rangle$ of the \planck\ sample filtered with the MF), but is unconstrained for RASS clusters. We find that the error is larger in the $V_x$ and $V_y$ directions due to the galactic cut. We also calculate the angle errors for a range of input velocities for four different input directions: the three perpendicular directions x ($l=0^{\circ}$, $b=0^{\circ}$), y ($l=90^{\circ}$, $b=0^{\circ}$) and z ($l=0^{\circ}$, $b=90^{\circ}$), as well as the reference $V_{\rm input}$ used above. The $95\%$ upper limit to the angle errors, $\Delta\alpha_{95}$, is shown in figure~\ref{f:chi95_angle}.

For maps filtered with the MF (UF) the $95\%$ confidence limit on the dipole direction for a 500 km/s input bulk flow is $\Delta\alpha_{95}^{500}=62^{\circ} (34^{\circ})$ for RASS clusters, $34^{\circ} (14^{\circ})$ for \planck\ clusters, and $9^{\circ} (4^{\circ})$ for EASS clusters. The errors are smaller than the discrepancies in the observed bulk flow directions. Therefore \planck\ should be able to better constrain the region where the bulk flow points to.

\subsection{Effect of the systematic error components on the amplitude of the recovered bulk flow}

The dominant systematic error is tSZ emission, and the dominant statistical error is CMB and instrument noise and uncertainty in the Y-parameter. They are velocity-independent and dominates for $v<1000$ km/s. At large velocities, the calibration error due to the scatter in optical depth dominates and gives a $<10\%$ error in the velocity.

Using inverse variance weights in the dipole fitting (as opposed to weighting by optical depth) is found to minimize the tSZ contribution. The recovered bulk flows are generally overestimated by $\approx20\%$ when $W=\tau/\sigma^2_{\rm noise}$ is used to weight pixels in the dipole fit.

\subsection{Performance of Filters}
The overall performance of both the MF and UF agree with expectations that maps filtered by the MF have higher kSZ signal to noise ratios while maps filtered by the UF have smaller thermal SZ bias. The UF is slightly a better filter in the \planck\ data since the dominant uncertainty of the measurement comes from the thermal SZ bias, and the recovered velocities are more accurately determined. Nevertheless, the thermal SZ bias is not completely removed by the UF. This is because we assume a $\beta$ model for the cluster profile, but assume a point source model to construct the filters. We could assume a $\beta$ model to construct the filters but we do not consider this here.

We now compare the filter performance in the \wmap\ channels which are studied in \osb\ (see their Figure 11) with the \planck\ channels in this work (Figure~\ref{f:vpplanck}), with the use of the RASS cluster sample which is used in both analysis. In \osb\ the MF was found to be more sensitive than the UF, but we find the opposite. The velocity detected at $95\%$ CL with the MF (UF) is reduced from $\sim5000$ km/s ($\sim10,000$ km/s) in the \wmap\ channels to $\sim1000$ km/s ($\sim1000$ km/s) in the \planck\ channels. Considering that the amplitude of observed bulk flow velocities are a few hundreds of km/s at scales $r<300 \mpch $ (Figure~\ref{f:dipole}), the filters in the \wmap\ channels do not have the sensitivity to measure bulk flow velocities at these scales. The differences between the \wmap\ and the \planck\ filters is expected: the error due to instrument noise is largely reduced because the noise levels of \planck\ are much lower than \wmap; the wider frequency coverage in the SZ sensitive regime also allows filters in the \planck\ channels to suppress the thermal SZ bias.

\section{Conclusion and Discussion}
\label{s:con}

We   investigated  \planck\  performances in determining bulk flow velocities using the kinetic Sunyaev-Zeldovich  effect. We characterize the sensitivity of the bulk flow velocity measurement using simulated \planck\ data combined with three representative all sky galaxy cluster surveys: the archived ROSAT All-Sky Survey, the \planck\ cluster catalog and the eRosita All-Sky Survey. We employ two different types of filters, a matched filter (MF) and an unbiased  kinetic SZ filter (UF), to maximize the cluster signal to noise ratio. 

The main results are (see also Table~\ref{t:summary}):
\begin{enumerate}

\item 
The use of simulated \planck\ sky maps instead of  \wmap\ ones in combination with the RASS catalog reduces the velocity that can be detected at $95\%$ CL,  from $\sim5000$ km/s ($\sim10,000$ km/s) to $\sim1000$ km/s ($\approx1000$ km/s) when filtered by the MF (UF).

\item Using all clusters with $z\le0.5$ we find that a bulk flow of 500 km/s would be measured in a strongly biased way with the RASS sample with both filters. The same velocity is measured with 0--5\% bias with EASS and Planck clusters, with uncertainties of 25--75 km/s. These numbers should be compared with the rms velocity out to $z=0.5$ in $\Lambda$CDM: 30 km/s (95\% upper limit: 48 km/s).

\item If the bulk flow is consistent with $\Lambda$CDM prediction of $v=30$ km/s at $z=0.5$, our analysis pipeline would obtain a 95\% upper limit to the recovered velocity (when the UF is used) of $v=470$ km/s for RASS cluster, $v=160$ km/s for \planck\ cluster sample and $v=60$ km/s for EASS cluster sample. This allows us to measure the departure from $\Lambda$CDM if the measured bulk flow is in excess of these estimates. 

\item \planck\ can also constrain a recovered bulk flow direction. For $V_{\rm input}=500$ km/s, the $95\%$ upper limit to the angle errors (when the UF is used) are $\Delta\alpha_{95}\approx30^{\circ}$ for the RASS clusters, $\Delta\alpha_{95}\approx15^{\circ}$ for the  \planck\ clusters, and $\Delta\alpha_{95}\approx5^{\circ}$ for the EASS clusters. These uncertainties are lower than the discrepancies observed in bulk flow directions at different optical depth up to now (Figure~\ref{f:dipole}). The errors in the directions contained within the galactic plane are larger than in the perpendicular one  due to the galactic cut at low latitudes.  

\item The error in the kinetic SZ dipole is dominated by the unfiltered CMB and instrument noise and intrinsic scatter of the Y-parameter, with a systematic thermal SZ bias. The contamination from extragalactic radio sources is negligible and can be safely ignored. The noise level is lowest for the EASS sample since it contains the most clusters.

\item The UF is more sensitive and effective than the MF in suppressing the bias induced by the thermal SZ in the velocity reconstruction. This is in contrast to the performance of the two filters on the \wmap\ maps, where the MF was much more sensitive than the UF. Differences in performances are to be expected, as the two experiments have different characteristics. As Planck has much lower noise, the tSZ contamination now plays a more major role in setting the bias and errors.~\planck\  increased frequency coverage and resolution, however,  enables the filters (UF in particular) to mitigate the effects of  the thermal SZ signal, especially when a large cluster sample is considered.
\end{enumerate}

We have not considered here the effect of infrared point sources, correlations in the signal  between CMB and clusters, large--scale (non bulk flow) peculiar velocities, and error determination in extracting optical depth from the survey in hand. Some of these could present further improvements to this initial study, which demonstrate the superior potentials \planck\  has in determining the bulk flow.

\begin{deluxetable}{cccccccc}
  \tablecaption{Summary of the results}
\tablewidth{0pc}
  \tablecolumns{8}
  \tabletypesize{\scriptsize}
  \tablehead{
  \colhead{} & \colhead{noise components} & \multicolumn{2}{c}{RASS} & \multicolumn{2}{c}{\planck} & \multicolumn{2}{c}{EASS}  \\
   \colhead{} & \colhead{} & \colhead{MF} & \colhead{UF} & \colhead{MF} & \colhead{UF} & \colhead{MF} & \colhead{UF}  \\
   }
  \startdata
  $V^{500}_{\rm rec}$ & noise + CMB + thermal SZ bias &$1085\pm82$ &$720\pm94$ & $530\pm126$& $516\pm73$ & $507\pm33$& $500\pm24$ \\
  $\left \langle v_x \right \rangle$ (75 km/s)& & $129\pm94$ & $131\pm138$ & $108\pm133$ & $88\pm68$ &$80\pm40$ & $75\pm29$ \\
  $\left \langle v_y \right \rangle$ (-426 km/s)& & $-75\pm105$ & $-434\pm133$ & $-432\pm123$ & $-431\pm73$ & $-427\pm37$& $-428\pm26$\\
  $\left \langle v_z \right \rangle$ (250 km/s)& & $1065\pm82$ & $526\pm66$ & $ 240\pm91$ & $255\pm49$ & $256\pm28$& $243\pm19$\\
   $\Delta\alpha^{500}_{95}$ & &$62^{\circ}$ & $34^{\circ}$ & $34^{\circ}$ & $14^{\circ}$  & $9^{\circ}$& $4^{\circ}$  \\  
   $\sigma_{\log V,\rm bin1}$ & & 0.71& 0.40& 0.12& 0.09 & 0.09& 0.09  \\
   $\sigma_{\log V,\rm bin2}$ & & 0.28& 0.14& 0.09& 0.04 & 0.02& 0.01 \\
   $\sigma_{\log V,\rm bin3}$ & & 0.08& 0.05& 0.04& 0.02 & 0.01& 0.01  \\
   $\sigma_{\log V,\rm bin4}$ & & 0.02& 0.03& 0.03& 0.02 & 0.01& 0.01  \\
   $\sigma_{\log V, \rm whole}$ & & 0.42& 0.24& 0.08& 0.06 & 0.05& 0.05 \\
\hline
   $V^{500}_{\rm rec} (\left \langle V_{\rm rec} \right \rangle)$ & only thermal SZ bias & $1100\pm60$ & $724\pm82$ & $530\pm115$ & $515\pm63$ & $511\pm30$& $504\pm14$ \\
    $\left \langle v_x \right \rangle$ (75 km/s)& & $125\pm73$ & $142\pm98$ & $82\pm109$ & $82\pm60$ & $104\pm30$ & $101\pm15$ \\
  $\left \langle v_y \right \rangle$ (-426 km/s)& & $-71\pm75$ & $-445\pm110$ & $-437\pm112$ & $-436\pm63$ & $-427\pm34$& $-428\pm14$ \\
  $\left \langle v_z \right \rangle$ (250 km/s)& & $1082\pm63$ & $534\pm46$ & $ 253\pm86$ & $250\pm46$ & $255\pm29$& $244\pm11$ \\
   $\Delta\alpha^{500}_{95}$ & &$62^{\circ}$ & $33^{\circ}$& $32^{\circ}$ & $14^{\circ}$  & $9^{\circ}$& $3^{\circ}$  \\  
   $\sigma_{\log V,\rm bin1}$ & & 0.68& 0.40& 0.12& 0.10 & 0.07& 0.04  \\
   $\sigma_{\log V,\rm bin2}$ & & 0.27& 0.15& 0.06& 0.03 & 0.03& 0.02  \\
   $\sigma_{\log V,\rm bin3}$ & & 0.09& 0.06& 0.04& 0.02 & 0.02& 0.01  \\
   $\sigma_{\log V,\rm bin4}$ & & 0.03& 0.03& 0.03& 0.02 & 0.01& 0.01  \\
   $\sigma_{\log V, \rm whole}$ & & 0.41& 0.24& 0.08& 0.06 & 0.05& 0.03 
    \enddata 
  \tablecomments{This table lists the main results of the bulk flow measurements for clusters within $z\le0.5$ in the three cluster surveys considered and for maps containing different components of uncertainties. The first five quantities are based on 100 realizations with input velocity of 500 km/s at $l=280^{\circ}$ and $b=30^{\circ}$. $V^{500}_{\rm rec}$ is the average value of the recovered velocity,  $\left \langle v_x \right \rangle$ , $\left \langle v_x \right \rangle$ , $\left \langle v_x \right \rangle$ are the average of the  individual directions. For ideal recovery, they should have values (75, -426, 250) km/s. $\Delta\alpha^{500}_{95}$ is the $95\%$ upper limit to the angle error. The last five quantities $\sigma_{\log V, \rm bin i}$ are the deviation parameters as defined in equation~\ref{eq:dispersionvv}, where bin i refers to the binned velocity range considered in the calculation: bin 1= 100--500 km/s, bin 2= 500--1000 km/s, bin 3= 1000--5000 km/s, bin 4= 5000--10000 km/s. The whole range is for velocities 100--10000 km/s. }
    \label{t:summary}
\end{deluxetable}

\section{Acknowledgements}
DSY Mak acknowledges support from the USC ProvostÕs Ph.D Fellowship Program. EP is an ADVANCE fellow (NSF grant AST-0649899). DSY Mak and EP acknowledge support from NASA grant NNX07AH59G and JPL Planck subcontract 1290790 and the warm hospitality of Aspen Center for Physics. SJO acknowledges useful discussions with Neelima Sehgal and support from the US Planck Project, which is funded by the NASA Science Mission Directorate. Some of the results in this paper have been derived using the HEALPix~\citep{Gorski2005} package. The authors acknowledge the use of the Planck Sky Model, developed by the Component Separation Working Group (WG2) of the Planck Collaboration. The authors thank the JPL data analysis group, Stefano Borgani and Sarah Church for fostering initial conversations on this topic.
 
 \newpage
\bibliography{planck_bf}
\appendix
\section{Derivation of Filter Kernel}
\label{a:filter}

\subsection{Data Model}
In real space, the observation field can be described as 
\begin{equation}
s_{\nu} (\theta)=y(0) (f_{\nu}-V) B_{\nu} (\theta) +n_{\nu} (\theta)
\end{equation}
\noindent where $y(0)$ is the y parameter at the cluster center (defined in equation B10), $f_{\nu} = x(e^x+1)/(e^x-1)-4$, and $x=(h\nu)/(k_{B}T_{\rm CMB})$ is the frequency response of the tSZ signal, $V=v_r m_e c/k_{\rm B}T_e$ gives the signal in kSZ effect, $B_\nu(\theta)$ is the convolved cluster profile given by:

\begin{eqnarray*}
 B_{\nu} (\theta) &=& \int{d{\Omega}' p({\theta}')b_{\nu}(\theta-\theta') } \\
		              &=& \sum^{\infty}_{l=0}{B_{l0,\nu}Y^0_l(\cos(\theta))}
\end{eqnarray*}

\noindent where $B_{l0,\nu}=\sqrt{(4\pi)/(2l+1)}b_{l0,\nu}p_{l0,\nu}$, $p(\theta)$ and $b_{\nu}(\theta)$ are the cluster spatial profile and beam function respectively, and $n_{\nu}$ is the noise consisting of instrumental noise and CMB . The noise map and the cross power spectrum $\mathbf{C}$ of each of the two components satisfy

\begin{eqnarray*}  
\left \langle  n_{lm,\nu_1}  n_{{l}'{m}',\nu_2}\right \rangle =\mathbf{C}_{l, \nu_1, \nu_2}\delta_{l{l}'} \delta_{m{m}'}
\end{eqnarray*}

\subsection{Filtered Field}
Let $\Phi_{\nu}$ and $u_{\nu}$ be the filter and filtered field respectively at frequency $\nu$, then 
\begin{equation}
u_{\nu}(\theta)=\int{d\theta^2\  s_{\nu}(\theta) \Phi_{\nu}(\theta) } \\
= \sum_{lm} u_{lm,\nu}Y_l^m(\beta)
\label{eq:ff}
 \end{equation}
\indent with $u_{lm,\nu} =\sqrt{\frac{4\pi}{2l+1}}s_{lm,\nu} \Phi_{l0,\nu}$

The total signal in all frequency channels is $u(\theta)=\sum_{\nu} u_{\nu} (\theta)$ 

A useful quantity to consider is the variance of the filtered map:
\begin{equation*}
\sigma^2_{u}=\left \langle  (u(\theta)-\left \langle  u(\theta) \right \rangle )^{2}\right \rangle 
\label{eq:var}
\end{equation*}

We introduce vector notation for the set of frequency dependent quantities, e.g. $\vec{B_l} = \left \{  B_{l,\nu_{1}},  B_{l,\nu_{2}}, \cdot B_{l,\nu_{N}}, \right \} $

Since $\left \langle u(\theta) \right \rangle = \left \langle \sum_{l} \left [ y_c (\vec{F_l} - V \vec{B_l} )+ \vec{n_l} \right ] \vec{\Phi_l} \right \rangle$, where we have used the notation $\vec{F_l}=\left \{  f_{\nu} B_{l0, \nu} \right \}_{\nu}$, then we have \\

\begin{eqnarray}
\sigma^2_{u} &= & \left \langle \left [ \sum_l{   (y_c \vec{F_l}-y_c V \vec{B_l})\vec{\Phi_l} + \vec{n_l}\vec{\Phi_l} - (y_c \vec{F_l}-y_c V \vec{B_l}) \vec{\Phi_l}   }  \right ]^{2} \right \rangle \nonumber \\
 &=& \left \langle \left [ \sum_l {\vec{n_l}\vec{\Phi_l} } \right ] ^{2} \right \rangle  \nonumber 
\end{eqnarray}

\noindent Therefore $\sigma^2_{u} = \sum_{l} { \vec{\Phi^{T}_{l} }\mathbf{C_l} \vec{\Phi_{l}} }$.

\subsection{Unbiased Matched Filter for kSZ signal}
In order to derive the optimal filter for the kSZ signal, we want to minimize $\sigma_u^2$ subject the the following constraints:
\begin{enumerate}
\item The filter is an unbiased estimator of KSZ signal at the source location, such that 
\begin{equation}
\sum_l { \vec{B_l} \vec {\Phi_{l}^{T} }} = 1 
\label{eq:c1}
\end{equation}

\item The filter should remove the TSZ signal at the source location, i.e. 
\begin{equation}
\sum_l { \vec{F_l} \vec {\Phi_{l}^{T}}} = 0
\label{eq:c2}
\end{equation}
\end{enumerate}

The functional variation of $\sigma_u^2$ with respect to $\vec{\Phi_l}$ subject to constraints~\ref{eq:c1} and~\ref{eq:c2} can be obtained using Lagrangian multipliers. The Lagrange function is defined as:

\begin{equation*}
L=\sum_{l} { \vec{\Phi^{T}_{l} }\mathbf{C_l} \vec{\Phi_{l}} } + \lambda_1 (1-\sum_l{  \vec{B_l} \vec{\Phi_l^{T}}} ) + \lambda_2 \sum_l{\vec{F_l} \vec{\Phi_l^{T}} }
\end{equation*}

Minimizing the Lagrange function with respect to the filter function $\vec{\Phi_l}$, 

\begin{eqnarray}
&& \Delta_{\vec{\Phi^{T}_l}} L = \sum_{l} { \left [ \mathbf{C_l} \vec{\Phi_l} - \lambda_1\vec{B_l} + \lambda_2 \vec{F_l} \right ] } = 0 \nonumber \\
&& \Rightarrow \mathbf{C_l} \vec{\Phi_l} = \lambda_1 \vec{B_l} - \lambda_2 \vec{F_l} \nonumber \\
&& \Rightarrow \vec{\Phi_l} = \mathbf{C_l^{-1}} (\lambda_1 \vec{B_l} - \lambda_2 \vec{F_l})
\label{eq:phi1} 
\end{eqnarray}

The job now is to find the constants $\lambda_1$ and $\lambda_2$. Using constraint~\ref{eq:c1} and~\ref{eq:phi1}, 

\begin{eqnarray}
&& \sum_l{\vec{B_{l}} \vec{\Phi_l^{T}} \vec{\Phi_l} }= \sum_l { \left [ \lambda_1 \vec{\Phi_l} \vec{B_l^{T}} \mathbf{C_l^{-1}} \vec{B_l} - \lambda_2 \vec{\Phi_l}\vec{B_l^{T}} \mathbf{C_l^{-1}} \vec{F_l}   \right ]} \nonumber \\
&& \Rightarrow 1 = \lambda_1 \sum_l{ \vec{B_l^{T}} \mathbf{C_l^{-1}} \vec{B_l}} -  \lambda_2 \sum_l{ \vec{B_l^{T}} \mathbf{C_l^{-1}} \vec{F_l}} \nonumber \\
&& \Rightarrow 1 = \lambda_1 \gamma - \lambda_2 \beta 
\label{eq:phi2}
\end{eqnarray}

Similarly, using constraint~\ref{eq:c2}  and~\ref{eq:phi1}, 
\begin{eqnarray}
&& \sum_l{\vec{F_{l}} \vec{\Phi_l^{T}} \vec{\Phi_l} }= \sum_l { \left [ \lambda_1 \vec{\Phi_l} \vec{F_l^{T}} \mathbf{C_l^{-1}} \vec{B_l} - \lambda_2 \vec{\Phi_l}\vec{F_l^{T}} \mathbf{C_l^{-1}} \vec{F_l}   \right ]} \nonumber \\
&& \Rightarrow 1 = \lambda_1 \sum_l{ \vec{F_l^{T}} \mathbf{C_l^{-1}} \vec{B_l}} -  \lambda_2 \sum_l{ \vec{F_l^{T}} \mathbf{C_l^{-1}} \vec{F_l}} \nonumber \\
&& \Rightarrow 1 = \lambda_1 \beta - \lambda_2 \alpha
\label{eq:phi3}
\end{eqnarray}

\noindent where $\alpha = \sum_l{\vec{F_l^T} \mathbf{C_l^{-1}} \vec{F_l} }$, $\beta=\sum_l{\vec{F_l^T} \mathbf{C_l^{-1}} \vec{B_l} }$, and $\gamma=\sum_l{\vec{B_l^T} \mathbf{C_l^{-1}} \vec{B_l}}$

Solving~\ref{eq:phi2} and~\ref{eq:phi3}, we obtain $\lambda_1 = \frac{\beta}{\alpha\gamma-\beta^2}$, and $\lambda_2=\frac{\alpha}{\alpha\gamma-\beta^2}$. Thus we 
get the filter kernal:
\begin{equation}
\vec{\Phi_l^{\rm UF}}  = \frac{\mathbf{C^{-1}_l}}{\alpha\gamma-\beta^2} (\alpha \vec{B_l}-\beta \vec{F_l})
\end{equation}
It is easy to verify that this filter kernel satisfies the two constraints.

The filter can be interpretated in this way:
\begin{enumerate}
\item $\Delta\equiv \alpha \gamma - \beta^2 $ is the normalization such that $\sum{\vec{B_l^{T}} \vec{\Phi_l} }= 1$
\item The term $\frac{\alpha}{\Delta} \vec{B_l}$ comes from the kSZ constraint, ensuring that the filter gives the kSZ signal at the source location.
\item The term $-\frac{\beta}{\Delta} \vec{F_l}$ comes from the TSZ constraint. Its purpose is to suppress the tSZ signal such that tSZ signal vanishes at the source location.
\item The filtered fields at each channel are weighted by the inverse of the covariance matrix $\mathbf{C_l^{-1}}$ and then combined to form the final filtered signal. 
\end{enumerate}

\subsection{Matched Filter}
The derivation of the matched filter is similar to that for the unbiased matched filter, except that  constraint~\ref{eq:c1} is used. Therefore we have the Lagrange function:
\begin{eqnarray}
&& L=\sum_{l} { \vec{\Phi^{T}_{l} }\mathbf{C_l} \vec{\Phi_{l}} } + \lambda (1-\sum_l{  \vec{B_l} \vec{\Phi_l^{T}}} )   \nonumber \\
&& \Rightarrow \Delta_{\vec{\Phi^{T}_l}} L = \sum_{l} { \left [ \mathbf{C_l} \vec{\Phi_l} - \lambda\vec{B_l}  \right ] } = 0 \nonumber \\
&& \Rightarrow \vec{\Phi_l} = \mathbf{C_l^{-1}} (\lambda \vec{B_l} )
\label{eq:mphi1} 
\end{eqnarray}

Using constraint~\ref{eq:c1} and solving for $\lambda$,  we find $\lambda=1/\gamma$. Thus we get the filter kernel:
\begin{equation}
\vec{\Phi_l^{\rm MF}}  = \frac{\mathbf{C^{-1}_l}}{\gamma} \vec{B_l}
\end{equation}

\section{Simulations of full-sky SZ maps}
\label{s:sim}

\subsection{Mass and Redshift Distribution of the cluster sample}
\label{ss:numcount}
The number of galaxy clusters per mass and redshift bin can be estimated using the comoving mass function:

\begin{equation*}
\frac{dN}{dMdz}(M,z) = \Delta \Omega\frac{dn}{dM} (M,z)\frac{dV}{dzd\Omega}(z)
\end{equation*}
\noindent where $\Delta\Omega$ is the solid angle in a given direction, and $\frac{dn}{dM}(M,z)$ is the mass function which describes the number density of galaxy clusters per mass bin at a given redshift z. Here we adopt the Jenkins mass function~\citep{Jenkins01} with the fitting formula given by:

\begin{eqnarray}
 \frac{dn(M,z)}{dM} = \frac{\rho_m}{M}\frac{d\ln \sigma^{-1}}{dM}f(\sigma^{-1}) \\
 f(\sigma^{-1}) = 0.315\exp(-\left | \ln \sigma^{-1} + 0.64 \right | ^{3.88} ) \nonumber
\end{eqnarray}
\noindent where $\rho_m$ is the present matter density, $\sigma$ is the variance of the mass fluctuation $\sigma_M^2 = \frac{1}{2\pi^2} \int^{\infty}_{0} k^2 dk P_m(k) \left | W_R(k) \right | ^2$, where $P_m(k)$ is the matter power spectrum $P(k,z)=A k^n T^2(k,z) D_+^2(z)$, 

We use the transfer function $T(k,z)$  from~\citet{Eisenstein98} that accounts for all baryonic effects in the matter transfer function on the large scale and an improved version of the BBKS transfer fitting formula. The growth function is defined as $D_+(z)=D_1(z) / D_1(0)$. We take the approximated form from~\citet{Lahav91} and~\citealt{Carroll92}: 

\begin{equation}
 D_1(z)= (1+z)^{-1} \frac{5\Omega_m(z)}{2} [ \Omega_m z^{4/7} - \Omega_{\Lambda}(z) + (1+\Omega_m z /2)(1+\Omega_{\Lambda}(z)/70) ]
 \label{eq:growth}
\end{equation}

By integrating over the mass and redshift interval of interest, we obtain the number of clusters N in a redshift and mass bin:

\begin{equation}
N_{\Delta M \Delta z} = 4\pi \int_{\Delta z} dz \int_{\Delta M} dM \frac{dN}{dM}{dz} (M,z)
\label{eq:numbercount}
\end{equation}

We then assign ${N}'$ clusters to a bin $(\Delta M, \Delta z)$ from a random Poisson distribution with an average value given by equation~\ref{eq:numbercount}. The mass range is taken to be $ 10^{14} -10^{16} M_\odot$ with a bin size of $\log (\Delta M)=0.1$. The minimum mass of cluster at redshift z that can enter the simulated catalog is given by $M_{\rm min}={\rm max}[10^{14}M_{\odot},M_{\rm lim}]$, where $M_{\rm lim}$ is the limiting mass of the corresponding survey from equation~\ref{eq:xr} and~\ref{eq:pla}. The redshift range z is $0-0.3$ with a bin size of $\Delta z =0.02$.  The number of clusters from this halo model is a strong function of $\sigma_8$ as shown in Figure~\ref{f:nofz}. 

Having obtained the mass and redshift distributions of the clusters, we distribute the clusters over the whole sky. For simplicity, we ignore the spatial correlation among clusters and assign a random position to each cluster.

\subsection{Modeling SZ signal of individual clusters}
\label{ss:scaling}
The SZ signal of galaxy clusters is characterized by the Compton y parameter, which in turn depends on the gas properties of individual clusters. The next step is to model the electron density $n_e(r)$, temperature $T_e(r)$ and the Integrated Compton Y parameter. 
\subsubsection{Gas properties from scaling relations}
Assuming that clusters are self-similar, virialized and isothermal, we obtain the mass-temperature relation from the virial theorem~\citep{Kaiser86}:

\begin{equation}
k_{\rm B} T_{\rm e} = \beta_{\rm T}^{-1}(1+z)\left ( \frac{\Omega_{\rm m} \Delta_{\rm V} (z)}{\Omega_{\rm m} (z)} \right )^{1/3} \left ( \frac{M_{\rm cl}}{10^{15} M_{\odot} h^{-1}} \right )^{2/3} \ {\rm keV}
\label{eq:mt}
\end{equation}

\noindent where $\beta_{\rm T}=0.75$ is the normalization constant under the assumption of hydrostatic equilibrium and isothermality, $\Delta_{\rm V}$ is the mean overdensity of a virialised sphere which we have calculated using the fitting formula from~\citet{Pierpaoli01}, $M_{\rm cl}$ is the mass of the cluster within the virial radius $r_{\rm vir}$, $M_{\rm cl}=4\pi/3\ \Delta_{\rm V}\rho_{\rm crit} r_{\rm vir}^3$. Solving for $r_{\rm vir}$ gives, 
\begin{equation}
r_{\rm vir}=\frac{9.5103}{1+z} \left ( \frac{\Omega_{\rm m} \Delta_{\rm V} (z)}{\Omega_{\rm m} (z)} \right )^{-1/3} \left ( \frac{M_{\rm cl}}{10^{15} M_{\odot} h^{-1}} \right )^{1/3} h^{-1} {\rm Mpc}
\label{eq:rvir}
\end{equation}
The assumption that clusters are isothermal is in good agreement with~\xmm\ observations of outer cluster regions which find that the cluster temperature profiles are isothermal within $\pm10\%$ up to $\approx r_{\rm vir}/2$~\citep{Arnaud04}.

The electron density within the virial radius described by a $\beta$ model (~\citealt{Waizmann09} and reference herein), 
\begin{equation}
n_{\rm e}(r) = n_{\rm e0} \left( 1+\frac{r^2}{r_c^2} \right ) ^{-3/2\beta}
\label{eq:ne}
\end{equation}
\noindent where $r_c$  is the core radius.~\citet{Geisbusch05} obtain a relationship between the core radius and the virial radius:
\begin{equation}
r_c(z) = 0.14 (1+z)^{1/5} r_{\rm vir}
\label{eq:rc}
\end{equation}
\noindent We choose $\beta=2/3$ to match~\citet{Waizmann09} which describes the X-ray surface brightness profile of observed clusters. 

The central electron density $n_{\rm e0}$ is normalized by $\int^{r_{\rm vir}}_{0} n_e(r) dV = N_e$, in which $N_e$ is the total number of electrons within the cluster given by:
\begin{equation}
N_e = \left ( \frac{1+f_{\rm H}}{2m_p} \right ) f_{\rm gas} M_{\rm cl}\
\label{eq:Ne}
\end{equation}
\noindent where $f_{\rm H}$ is the hydrogen fraction, $f_{\rm gas}=\Omega_{\rm b} / \Omega_{\rm m}$ is the baryonic gas mass fraction of the total cluster mass. Here we take $f_{\rm gas} = 0.168$ from WMAP 5 year results and $f_{\rm H}=0.76$. We find,
\begin{equation}
n_{\rm e0} = \frac{N_e}{4\pi r_{\rm c}^3 [ \frac{1}{(0.14(1+z)^{1/5})} + \tan^{-1}(0.14(1+z)^{1/5}) - \frac{\pi}{2}}
\label{eq:neo}
\end{equation}

The Comptonization parameter is proportional to the product of the electron temperature and density, and is given by $y=\int dl (k_{\rm B}T_e)/(m_e c^2) n_e \sigma_{\rm T}$. Performing this integral along the line-of-sight of the cluster, we obtain:

\begin{equation}
y(\theta)=2\frac{k_{\rm B} T_{\rm e}}{m_{\rm e} c^2} \frac{\sigma_{\rm T} r_{\rm c} n_{\rm e0}}{\sqrt{1+\frac{\theta^2}{\theta_c^2}}} \tan^{-1} \sqrt{ \frac{\frac{1}{(0.14(1+z)^{2}} -\frac{\theta^2}{\theta_c^2}}{1+\frac{\theta^2}{\theta_c^2}}}
\label{eq:comp}
\end{equation}

\noindent where $\theta_c=r_c/D_{\rm A}$. The SZ signal is then calculated using this compton y parameter for each cluster within the virial radius. It is easy to show from equation B10 that y drops to zero when $\theta \geq \theta_{\rm vir}$. 
The analytic total Comptonization parameter $Y$ is obtained simply by integration equation B10 over the cluster size:

\begin{equation}
Y=\int d\Omega y(\theta) = D_{\rm A}^{-2}(z)\frac{k_{\rm B}\sigma_{\rm T}}{m_{\rm e} c^2} \int dV n_{\rm e} T_{\rm e} = \frac{k_{\rm B}T_{\rm e}\sigma_{\rm T} N_{\rm e}}{m_{\rm e} c^2 D_{\rm A}^2}
\label{eq:icomp}
\end{equation}
\noindent where $D_{\rm A}$ is the angular diameter distance to the cluster. We then scale the pixels within the disc of angular size $\theta_{\rm vir}$ by a factor $N=\frac{Y}{\sum_i{y_i d\Omega_{\rm pix}}}$, where $y_i$ is the y value in the center of pixel $i$ and $d \Omega_{\rm pix}$ is the pixel solid angle. 

There is an expected cluster mass--optical depth scatter in nature due to the fact that unknown cluster physics can affect the exact form and normalization of the scaling relations, e.g. $Y-M$ relation.~\citet{Sehgal2007} have estimated the typical intrinsic scatter in the $Y-M$ relation with simulations of the ACT clusters and found that this is about $15\%$ in Y. While ACT clusters represented in those simulations have masses that are lower than those expected in our samples (in particularly \planck), we do not expect bigger objects which are more virialized to have a larger scatter. We therefore introduce a 15\% scatter in the $Y$ parameter in our simulations for all cluster samples. In principle, there is also a measurement error for optical depth due to inaccuracy of SZ photometry. However, we ignore this effect because such uncertainty is not yet well characterized for \planck.

\end{document}